\documentclass[10pt]{article} 
\usepackage[accepted]{tmlr_arxiv}
\usepackage{versions}
\usepackage{xspace}
\usepackage[inline]{enumitem}
\setlist{nolistsep}
\usepackage{wrapfig}
\usepackage{adjustbox}
\usepackage{multirow}
\usepackage{graphics}
\PassOptionsToPackage{hyphens}{url}\usepackage[colorlinks=true]{hyperref}
\usepackage{amsfonts}
\usepackage{floatflt}
\usepackage{amsmath}
\usepackage{array}
\usepackage[T1]{fontenc}
\usepackage{booktabs}
\usepackage{colortbl}
\usepackage{soul}
\usepackage[hang,flushmargin]{footmisc}
\newcommand{\Snospace}{\S{}}

\hypersetup{citecolor=blue,linkcolor=black,urlcolor=black}

\usepackage{tikz}
\usetikzlibrary{shapes,arrows}
\usetikzlibrary{backgrounds, positioning, fit}
\usepackage{array}
\newcolumntype{C}[1]{>{\raggedright\arraybackslash}p{#1}}

\usepackage{wasysym}
\usepackage{pifont}
\usepackage{amsmath}

\definecolor{ForestGreen}{RGB}{34,139,34}
\newcommand{\smark}{$\LEFTcircle$\xspace}
\newcommand{\xmark}{$\Circle$\xspace}
\newcommand{\cmark}{$\CIRCLE$\xspace}
\newcommand{\graymark}{{\color{black}$\Xi$}}
\newcommand{\redmark}{{\color{red}$\Xi$}}
\newcommand{\greenmark}{{\color{ForestGreen}$\Xi$}}

\newcommand{\nocolorNaive}{{$\Psi$}}
\newcommand{\redmarkNaive}{{\color{red}$\Psi$}}
\newcommand{\greenmarkNaive}{{\color{ForestGreen}$\Psi$}}

\newcommand{\nocolorOurs}{{$\Delta$}}
\newcommand{\redmarkOurs}{{\color{red}$\Delta$}}
\newcommand{\greenmarkOurs}{{\color{ForestGreen}$\Delta$}}

\newcommand{\dtrain}{\mathcal{D}_{tr}\xspace}
\newcommand{\dtest}{\mathcal{D}_{te}\xspace}

\newcommand{\model}{f\xspace}
\newcommand{\adv}{\protect{$\mathcal{A}dv$}\xspace}

\newcommand{\defevasion}{$\textsc{EvsnRob}$\xspace}
\newcommand{\defevasionPre}{$\textsc{EvsnRob.Pre}$\xspace}
\newcommand{\defevasionIn}{$\textsc{EvsnRob.In}$\xspace}
\newcommand{\defevasionPost}{$\textsc{EvsnRob.Post}$\xspace}

\newcommand{\defoutrem}{$\textsc{PoisnRob}$\xspace}
\newcommand{\defoutremPre}{$\textsc{PoisnRob.Pre}$\xspace}
\newcommand{\defoutremIn}{$\textsc{PoisnRob.In}$\xspace}
\newcommand{\defoutremPost}{$\textsc{PoisnRob.Post}$\xspace}

\newcommand{\defdiffpriv}{$\textsc{DiffPriv}$\xspace}
\newcommand{\defdiffprivPre}{$\textsc{DiffPriv.Pre}$\xspace}
\newcommand{\defdiffprivIn}{$\textsc{DiffPriv.In}$\xspace}
\newcommand{\defdiffprivPost}{$\textsc{DiffPriv.Post}$\xspace}

\newcommand{\defmodelwm}{$\textsc{MdlWM}$\xspace}
\newcommand{\defmodelwmPre}{$\textsc{MdlWM.Pre}$\xspace}
\newcommand{\defmodelwmIn}{$\textsc{MdlWM.In}$\xspace}
\newcommand{\defmodelwmPost}{$\textsc{MdlWM.Post}$\xspace}

\newcommand{\defdatawm}{$\textsc{DtWM}$\xspace}
\newcommand{\defdatawmPre}{$\textsc{DtWM.Pre}$\xspace}

\newcommand{\deffngprnt}{$\textsc{Fngrprnt}$\xspace}
\newcommand{\deffngprntPost}{$\textsc{Fngrprnt.Post}$\xspace}

\newcommand{\deffair}{$\textsc{GpFair}$\xspace}
\newcommand{\deffairPre}{$\textsc{GpFair.Pre}$\xspace}
\newcommand{\deffairIn}{$\textsc{GpFair.In}$\xspace}
\newcommand{\deffairPost}{$\textsc{GpFair.Post}$\xspace}

\newcommand{\defexpl}{$\textsc{Expl}$\xspace}
\newcommand{\defexplPost}{$\textsc{Expl.Post}$\xspace}

\newcommand{\defense}{$\textbf{\texttt{D}}$\xspace}

\newcommand{\defenseone}{$\textbf{\texttt{D}}_1$\xspace}
\newcommand{\defensetwo}{$\textbf{\texttt{D}}_2$\xspace}
\newcommand{\defensethree}{$\textbf{\texttt{D}}_3$\xspace}

\newcommand{\risk}{$\textbf{\texttt{Rk}}$\xspace}
\newcommand{\riskone}{$\textbf{\texttt{Rk}}_1$\xspace}
\newcommand{\risktwo}{$\textbf{\texttt{Rk}}_2$\xspace}

\newcommand{\phiacc}{$\phi_{\text{u}}$\xspace}

\newcommand{\upmark}{{$\wedge$}}
\newcommand{\downmark}{{$\vee$}}
\newcommand{\samemark}{{$\sim$}}

\newcommand{\method}{$\textsc{Def\textbackslash Con}$\xspace}

\newcommand\change[1]{{#1\xspace}}

\usepackage{amssymb,amsmath,amsthm}
\usepackage{thmtools}
\usepackage[most]{tcolorbox}
\definecolor{White}{rgb}{1, 1, 1}
\definecolor{Periwinkle}{rgb}{0, 0, 0}
\colorlet{LightGray}{White!98!Periwinkle}
\definecolor{blond}{rgb}{0.98, 0.94, 0.75}
\definecolor{bubblegum}{rgb}{0.99, 0.76, 0.8}
\definecolor{mossgreen}{rgb}{0.68, 0.87, 0.68}
\definecolor{teagreen}{rgb}{0.82, 0.94, 0.75}

\declaretheoremstyle[
    numbered=no, 
    headfont=\bfseries,
    postheadspace=0pt,
    headpunct={}
]{thmsty}

\declaretheorem[style=thmsty, name={}]{guidelines}
\declaretheorem[style=thmsty, name={}]{takeaway}

\tcolorboxenvironment{guidelines}{enhanced jigsaw, colback=LightGray, drop shadow, boxrule=0.9pt, boxsep=0.1pt, left=4pt, right=4pt, top=4pt, bottom=4pt}

\tcolorboxenvironment{takeaway}{enhanced jigsaw, colback=blond, drop shadow, boxrule=0.9pt, boxsep=0.1pt, left=4pt, right=4pt, top=4pt, bottom=4pt}

\includeversion{submit}
\excludeversion{arxiv}

\title{Combining Machine Learning Defenses without Conflicts}

\author{\name Vasisht Duddu \email vasisht.duddu@uwaterloo.ca \\
      \addr University of Waterloo
      \AND
      \name Rui Zhang \email zhangrui98@zju.edu.cn \\
      \addr Zhejiang University
      \AND
      \name N. Asokan \email asokan@acm.org\\
      \addr University of Waterloo}

\def\openreview{\url{https://openreview.net/forum?id=C7FgsjfFRC}} 

\begin{document}

\maketitle

\begin{abstract}
Machine learning (ML) models require protection against various risks to security, privacy, and fairness. 
Real-life ML models need simultaneous protection against multiple risks,  necessitating combining multiple defenses \emph{effectively}, without incurring significant drop in the effectiveness of the constituent defenses.
We present a systematization of existing work based on \emph{how defenses are combined}, and \emph{how they interact}.
We then identify unexplored combinations, and evaluate combination techniques to identify their limitations.
Using these insights, we present, \method, a combination technique which is (a) \emph{accurate} (correctly identifies whether a combination is effective or not), (b) \emph{scalable} (allows combining multiple defenses), (c) \emph{non-invasive} (allows combining existing defenses without modification), and (d) \emph{general} (is applicable to different types of defenses).
We show that \method achieves 90\% accuracy on eight combinations from prior work, and 86\% in 30 unexplored combinations which we empirically evaluated.
\end{abstract}

\section{Introduction}\label{sec:introduction}

Machine learning (ML) models are susceptible to a wide range of risks to security~\citep{SoKMLPrivSec,poisonSurvey}, privacy~\citep{privacySurvey,miaSurvey}, and fairness~\citep{fairSurvey,fairsurvey2}. Several defenses have been proposed to mitigate them.
Real-life models require simultaneous protection against multiple risks.
But defenses designed to protect against one risk~\citep{certifiedRobustness2020sok,robustSoK,privacySurvey,fairSurvey} may impact susceptibility to other unrelated risks~\citep{duddu2023sok}.
This raises the question of how to combine defenses \emph{effectively}, without incurring a significant drop in the protection provided by each constituent defense.
Practitioners need effective \emph{combination techniques}, either by \emph{modifying existing defenses} (invasive), or by identifying whether \emph{existing defenses can be combined without modification} (non-invasive). 
Prior work is either limited to specific defenses (e.g.,~\citet{szyller2023conflicting,fedtradeoff,fioretto2022differential,sokExpl}), or study interactions among defenses with risks~\citep{duddu2023sok,surveyTradeoff}. 
No prior work provides a way to quickly determine if  \emph{defenses can be combined effectively}.

We first systematically survey existing work on combining defenses based on: 
\begin{enumerate*}[label={(\alph*)}]
\item \emph{how defenses are combined} (i.e., what combination technique was used), and
\item \emph{how they interact} (i.e., whether they conflict or align).
\end{enumerate*}
We then identify previously unexplored combinations, and evaluate prior combination techniques to identify their limitations.

Non-invasive combination techniques are easier to deploy as they do not require expert knowledge from practitioners. Therefore,
we identify ``\emph{mutually exclusive placement}''~\citep{yaghini2023learning} as a promising technique.
It presumes that two defenses can be effectively combined iff they \emph{operate on different stages in the ML pipeline}: pre-, in-, and post-training. 
However, it can result in ineffective combinations: \begin{enumerate*}[label={(\alph*)}]
\item later-stage defenses can conflict with earlier ones (e.g., model or dataset watermarking with adversarial training or differential privacy~\citep{szyller2023conflicting}), and \item same-stage defenses may still be compatible (see \Snospace\ref{sec:evaluation}).
\end{enumerate*}

Based on these insights, we present \method, a technique which is 
\begin{enumerate*}[label=(\roman*)]
\item \emph{accurate} (correctly identifies whether a combination is effective or not),
\item \emph{scalable} (allows two or more defenses to be combined),
\item \emph{non-invasive} (does not require changes to the defenses being combined), and
\item \emph{general} (applicable to different types of defenses).
\end{enumerate*}
\method is inspired by mutually exclusive placement (aka na\"ive technique), but overcomes its limitations by explicitly addressing the reasons that underlie conflicts among defenses: a later-stage defense either \begin{enumerate*}[label={(\alph*)}]
\item mitigates a risk re-purposed as a defense by an early-stage defense, or 
\item overrides changes made by an early-stage defense~\citep{szyller2023conflicting}.
\end{enumerate*}
We claim the following contributions:
\begin{enumerate}[leftmargin=*]
\item a \emph{systematization} of prior work based on  combination techniques, and types of resulting interactions; (\Snospace\ref{sec:systematization})

\item \emph{identifying} unexplored combinations, and \emph{evaluating} prior techniques for limitations.; (\Snospace\ref{sec:unexplored} and \Snospace\ref{sec:desiderata})

\item \method\footnote{Link to Code: \url{https://github.com/ssg-research/combining-defenses}.}, a scalable, non-invasive, and general combination technique (\Snospace\ref{sec:approach}), which is more accurate than the na\"ive technique, with a balanced accuracy of
\begin{itemize}[leftmargin=*]
    \item 90\% (\method) vs. 40\% (na\"ive) using eight combinations from prior work as ground truth, (\Snospace\ref{sec:evalPriorWork}),
    \item 86\% (\method) vs. 36\% (na\"ive) via empirical evaluation of 30 unexplored combinations (\Snospace\ref{sec:evalEmpirical} and~\ref{sec:hyperparam}).
\end{itemize}
\method constitutes an inexpensive and fast technique for practitioners to determine if a combination of existing defenses without modification can be effective, without using expensive empirical evaluation.
\end{enumerate}
\section{Background: ML Notations}\label{sec:background}

Consider $X$ as the space of all possible input data records (e.g., images, text prompts) and $Y$ as the space of corresponding outputs (e.g., classification labels for classifiers, predicted next tokens for generative models).
An ML model is a function $\model^{\theta}$ which maps $x$ to $y$, i.e., $\model^{\theta}: X \rightarrow Y$ where $\theta$ indicates the model's parameters.
Hereafter, we denote $\model^\theta$ by simply writing $\model$.
We consider two datasets of the form ($x$, $y$), where $x$ is the input data record and $y$ is the output, for training an ML model using training dataset ($\dtrain$) and evaluate the model on test dataset ($\dtest$).
We focus our evaluation (\Snospace\ref{sec:evaluation}) on classifier models, and describe the training and inference for classifiers. 

\noindent\textbf{Training.} We iteratively update $\theta$ using ($x, y$) $\in$ $\dtrain$ over multiple epochs to minimize some objective function $\mathcal{C}$: $\min\limits_{\theta} \, l(\model(x),y;\theta) + \lambda \, R(\theta)$ where $l(\model(x),y)$ is the prediction error on $x$ for the ground truth $y$.
$R(\theta)$ is the regularization function which restricts $\theta$ from taking large values and $\lambda$ is a hyperparameter to control regularization.
The parameters are updated as: $\theta:=\theta - \alpha \frac{\partial \mathcal{C}}{\partial \theta}$ where $\alpha$ is the learning rate.

\noindent\textbf{Inference.} We measure the utility of $\model$ using its accuracy on $\dtest$ computed as 
\begin{equation*}
\phi_{u}(\model, \mathcal{D}_{\text{test}}) 
= \frac{1}{|\mathcal{D}_{\text{test}}|} 
\sum_{(x, y) \in \mathcal{D}_{\text{test}}} 
\mathbb{I} \left\{ \hat{\model}(x) = y \right\}
\end{equation*}
where $\hat{\model}(x)$ is the most likely class.
If \phiacc is acceptable, $\model$ is deployed to provide predictions for input $x$, represented by $\model(x)$ for the probability vector across different classes.

\section{Framework for Systematization}\label{sec:framework}

Given a list of defenses whose combinations are explored (\Snospace\ref{sec:defenses}), we present a framework to systematize prior work based on \emph{how the defenses are combined} (\Snospace\ref{sec:combinationTechnique}), and \emph{how they interact} (\Snospace\ref{sec:factorsInteractions}). We then discuss the completeness of our framework and how to extend it (\Snospace\ref{sec:completeness}).

\subsection{Defenses being Combined}\label{sec:defenses}

We describe various defenses proposed to mitigate risks to ML models in the presence of an adversary (\adv). As part of our systematization (\Snospace\ref{sec:systematization}), we will later enumerate all possible pairwise combination of these defenses.

\noindent\textbf{\underline{Evasion robustness (\defevasion})}  protects against \emph{evasion}, which forces $\model$ to misclassify an input $x$ by adding perturbation $\delta_{rob}$ (aka adversarial examples)~\citep{robustSoK,madry2018towards}. 
Here, $\delta_{rob} = argmax_{\delta_{rob}} l(f(x+\delta_{rob},y)$ and $||\delta_{rob}||<\epsilon_{rob}$, where $\epsilon_{rob}$ is a perturbation budget.

\noindent\textbf{\underline{Poison robustness (\defoutrem})} protects against \emph{poisoning} which involves training $\model$ on \emph{poisons} which are obtained by either tampering existing data records in $\dtrain$ or adding manipulated data records to $\dtrain$ to degrade \phiacc~\citep{poisonSurvey}. Alternatively, poisoning for backdoors forces $\model$ to incorrectly learn a mapping of some pattern in the poisons, to a target class chosen by \adv. 
During inference, any data record with that pattern is then misclassified to the target class~\citep{li2022backdoor}.

\noindent\textbf{\underline{Model Watermarking (\defmodelwm})} checks for \emph{unauthorized model ownership}, including \emph{model extraction attacks} where \adv trains a local \emph{surrogate model} to mimic the functionality of $\model$~\citep{orekondy2019knockoff}.
\defmodelwm embeds watermarks in $\model$ that transfer to the surrogate model during extraction.
If a suspect model's watermark accuracy is above some pre-defined threshold, it is identified as a surrogate.

\noindent\textbf{\underline{Fingerprinting (\deffngprnt})}  also checks for \emph{unauthorized model ownership} by generating unique identifiers or \emph{fingerprints} (e.g., adversarial examples, embeddings), for $\model$.
These fingerprints transfer from $\model$ to any surrogate model that are derived from it but are distinct from the fingerprints of independently trained models~\citep{caoFingerprint,peng2022fingerprinting,lukas2021deep,paramfingerprint,maini2021dataset}.

\noindent\textbf{\underline{Data watermarking (\defdatawm})}  checks for \emph{unauthorized data use} where $\model$ is trained on datasets collected without consent (e.g., face images for facial recognition)~\citep{sablayrolles2020radioactive,huang2021unlearnable,wenger2023sok}.
\defdatawm either augments $\dtrain$ with watermarks (e.g., backdoors)~\citep{tekgul2022effectiveness,sablayrolles2020radioactive}, or selects high-influence samples from $\dtrain$ as watermarks~\citep{liu2022miafingerprint}.
For verification, we check whether watermarks were in $\dtrain$ using statistical tests~\citep{sablayrolles2020radioactive} or membership inference~\citep{liu2022miafingerprint}.
The difference between \defmodelwm and \defdatawm is how a model trained from scratch on $\dtrain$ is classified: \defdatawm flags it for unauthorized data use while \defmodelwm classifies it as independently trained.

\noindent\textbf{\underline{Differential privacy (\defdiffpriv})} protects against membership inference (whether a data record was in $\dtrain$)~\citep{miaSurvey} and data reconstruction (reconstructing data records in $\dtrain$)~\citep{modelInv} by hiding whether an individual's data record was used to train $\model$~\citep{abadi2016deep}. 
Given two models trained on neighboring datasets differing by one record, \defdiffpriv bounds the privacy loss (distinguishability in predictions between the two models) by $e^\epsilon_{dp} + \delta$. Here, $e^\epsilon_{dp}$ is the privacy budget and $\delta_{dp}$ is probability where the privacy loss is $>e^\epsilon_{dp}$.

\noindent\textbf{\underline{Group fairness (\deffair})}  minimizes \emph{discriminatory behavior} to ensure equitable behavior across demographic groups identified by a sensitive attribute in $x$ (e.g., race or sex)~\citep{fairSurvey,fairsurvey2}. 
\deffair is measured using various metrics like accuracy parity, demographic parity~\citep{zafarfairness}, equalized odds and equality of opportunity~\citep{hardt2016equality}.

\noindent\textbf{\underline{Explanations (\defexpl})}  give insights into $\model$'s \emph{incomprehensible behavior}~\citep{transparencySurvey} which can be used to detect \emph{discriminatory behavior}~\citep{gradcam,kim2018interpretability}. 
Explanations $\gamma(x)$ indicate the influence of different input attributes in $x$ on $\model(x)$.
There are three main categories: \emph{Attribution-based}~\citep{ismail2021improving,Smilkov2017SmoothGradRN,intgrad}; \emph{influence-based}~\citep{koh17a}; and \emph{recourse-based}~\citep{Wachter2017CounterfactualEW}.
We focus on attribution-based explanations which are popular in prior work on combining defenses, and applicable to various domains (e.g., tabular, image).
These explanations require training a linear model in a region around a point of interest $x$~\citep{ismail2021improving,Smilkov2017SmoothGradRN,intgrad}.
The coefficients of the linear model for an input $x = (x_1, \cdots x_n)$ with $n$ attributes, constitutes $\gamma(x)$.

We summarize the defenses and their impact on \phiacc in Table~\ref{tab:defenses}.

\subsection{Combination Techniques}\label{sec:combinationTechnique}

Based on our survey described later in \Snospace\ref{sec:systematization}, we identify two combination techniques which either \emph{modify existing defenses}, or identify whether \emph{existing defenses can be combined without modification}. 
We mark them as \ref{t1:optimization} and \ref{t2:intervention} respectively, and describe them as follows:
\begin{enumerate}[label=\textbf{T\arabic*},leftmargin=*,leftmargin=*,wide, labelindent=0pt]
\setlength\itemsep{0em}
\item \label{t1:optimization} \textbf{(Optimization)} includes game-theoretic formalization, regularization, or constrained equation solving. \ref{t1:optimization} incorporates defenses into the objective function (e.g., regularization terms) so that the corresponding defense constraints can be satisfied during training for an effective combination~\citep{xin2023connection,hu2023outlier,bu2022practical,wu2023augment,zhang2022differentially,he2020robustness,tran2022fairness,farmur,benz2021robustness,MaNeurips2022,nanda2021fairness,fairOrRobbust,SunFairRobust,li2023wat,lee2024dafa,wei2023cfa,failof,liu2021fairness,shekhar2021fairod,zhang2021towards,tran2021lagrange,LIU2022108,lowy2023stochastic,dpfair,tran2021differentially,yaghini2024regulation,ding2020differentially,DPLogistic,zhang2021balancing,esipova2023disparate,xu2020removing,tran2023fairdp,lakkaraju2020robust,NEURIPS2019_172ef5a9,Li_2023_CVPR}.
This also includes using variants of standard model architectures and algorithms, specifically catered for a particular combination to give better trade-offs among the defenses~\citep{ding2020differentially,DPLogistic,yang2022differentially,phan2019heterogeneous,phan2020scalable}.
\item \label{t2:intervention} \textbf{(Mutually Exclusive Placement)} consists of applying defenses at \emph{different} stages of the ML pipeline---
\begin{enumerate*}[label=\roman*),itemjoin={,\xspace}]
\item pre-training (modifies $\dtrain$)
\item in-training (modifies training configuration such as objective function)
\item post-training (modifies inputs or outputs of trained $\model$ during inference)
\end{enumerate*}---to avoid conflicts~\citep{yaghini2023learning,DPExplanations}. 
We later refer to \ref{t2:intervention} as the \emph{na\"ive technique} and use it as a baseline to compare with our proposed technique \method (\Snospace\ref{sec:approach} and \Snospace\ref{sec:evaluation}).
\end{enumerate}

\subsection{Type of Interactions}\label{sec:factorsInteractions}
Consider two defenses \defenseone and \defensetwo which protect against risks \riskone and \risktwo respectively, with \defensetwo is applied after \defenseone.
The can interact in one of two ways:
\begin{itemize}[leftmargin=*]
\item \textbf{Alignment.} \defenseone and \defensetwo are \emph{aligned} if any of these hold:
\begin{enumerate*}[label=(\roman*),itemjoin={;\xspace}]
\item \defenseone and \defensetwo  do not impact \risktwo and \riskone, respectively
\item \defenseone reduces \risktwo, increasing \defensetwo's effectiveness
\item \defenseone generalizes \defensetwo, so its effectiveness implies that of \defensetwo.
\end{enumerate*}
Alignment leads to an effective defense combination.
When one defense implies the other (case (iii)), applying the first may be sufficient since we get the second for free (e.g., attribute privacy and group fairness~\citep{aalmoes2022alignment}).

\item  \textbf{Conflict.} \defenseone and \defensetwo \emph{conflict} if any of the following hold:
\begin{enumerate*}[label=(\roman*),itemjoin={;\xspace}]
\item \defenseone uses risk \risk (protected by \defensetwo), making \defenseone ineffective
\item \defensetwo overrides \defenseone's changes, making \defenseone ineffective.
\end{enumerate*}
Conflict leads to an ineffective combination of the defenses.
To avoid conflicts, we need accurate combination techniques.
\end{itemize}

\subsection{Completeness of Framework}\label{sec:completeness} 

In \Snospace\ref{sec:defenses}, we identify some ML defenses for analysis. We do not claim that this list is complete. 
For instance, there are other defenses (e.g., \emph{individual fairness}~\citep{dworkindiv}, and \emph{interpretability}~\citep{kleinberg2019simplicity}), or defenses specific to models other than classifiers (e.g., language and diffusion models), and settings (e.g., federated learning). 
We can update the framework (\Snospace\ref{sec:defenses}) to add new defenses and enumerate all its combinations with other defenses (as shown later in Table~\ref{tab:interactions}).
Similarly, in \Snospace\ref{sec:combinationTechnique}, we do not claim that the list of combination techniques is complete, but it covers all the techniques seen in our systematization (\Snospace\ref{sec:systematization}).
New combination techniques can be easily added into our framework, and later used for systematization as shown in \Snospace\ref{sec:methodology}.
\section{Systematizing Interactions among Defenses}\label{sec:systematization}

\noindent We now use our framework to categorize existing literature. We present our methodology (\Snospace\ref{sec:methodology}), and show the systematization of prior work (\Snospace\ref{sec:survey}).

\subsection{Methodology}\label{sec:methodology}

We enumerate all defense combinations from \Snospace\ref{sec:defenses}, in Table~\ref{tab:interactions}. 
Each combination is represented as a cell in Table~\ref{tab:interactions}, for which we indicate related work, combination technique used, and the type of resulting interaction.

\begin{guidelines}
For each prior work, we identify the following:\\
\textbf{Combination Technique}: We mark the technique to combine defenses as \ref{t1:optimization} or \ref{t2:intervention}. \\
\textbf{Type of Interaction}: We mark the interactions as a conflict (\redmark), alignment (\greenmark), or unexplored (\graymark).
\end{guidelines}

\noindent\textbf{Justification.}
Prior surveys are limited to specific defenses (e.g., ~\citet{fedtradeoff,fioretto2022differential,sokExpl}), or do not cover sufficient details to about combination techniques (e.g., \citet{surveyTradeoff,ferry:sok}).
This makes it challenging to design better combination techniques, and systematically compare with prior works.
Our systematization addresses these limitations by (a) covering multiple defenses and their combinations, and (b) explicitly mapping them to the combination techniques and the type of resulting interactions.
As shown later in \Snospace\ref{sec:application}, our systematization helps to identify gaps in existing literature (e.g., unexplored combinations, and limitations of prior techniques).
Using the insights from our systematization, we can design and evaluate a new combination technique (\Snospace\ref{sec:approach} and~\Snospace\ref{sec:evaluation}).

\noindent\textbf{Selecting Papers for Analysis.} We started surveying papers in Google Scholar using keywords (e.g, ``combining <defense 1> and <defense 2>''). We selected all papers including those published in top-tier ML and security/privacy venues (e.g., NeurIPS, ICML, ICLR, AAAI, CCS, S\&P), related workshop papers, and unpublished papers on ArXiv.
We examined their citations and related work to find other papers. 
Finally, we used papers from related surveys (e.g., \citet{surveyTradeoff,fedtradeoff,fioretto2022differential,sokExpl,ferry:sok}) to ensure a comprehensive coverage.

\setlength\tabcolsep{2pt}
\begin{table*}[!htb]
\caption{\textbf{Overview of Pairwise Combinations among Defenses}: For each combination cell, we citep related work and indicate the ``interaction type'' (\greenmark\xspace $\rightarrow$ alignment, \redmark\xspace $\rightarrow$ conflict, \graymark\xspace $\rightarrow$ unexplored), and ``combination technique'' used (\ref{t1:optimization}-\ref{t2:intervention}).}
\centering
\scriptsize
\resizebox{\textwidth}{!}{ 
\begin{tabular}{l | C{2.3cm} | C{2.2cm} | C{1.9cm} | C{1.9cm} | C{1.9cm} | C{2.2cm} | C{1.9cm} }
\bottomrule

\toprule
 & \textbf{\defevasion} & \textbf{\defoutrem} & \textbf{\defmodelwm} & \textbf{\deffngprnt} & \textbf{\defdatawm} & \textbf{\defdiffpriv} & \textbf{\deffair} \\
\midrule
\multirow{2}{*}{\textbf{\defoutrem}}  & \greenmark $\rightarrow$ \ref{t1:optimization}: \citep{xin2023connection,hu2023outlier} & \cellcolor{black!30}  & \cellcolor{black!30} & \cellcolor{black!30} & \cellcolor{black!30} & \cellcolor{black!30} & \cellcolor{black!30} \\

\midrule

\multirow{4}{*}{\textbf{\defmodelwm}} & \redmark $\rightarrow$ \ref{t2:intervention}: \citep{szyller2023conflicting}  & \graymark &  \cellcolor{black!30} & \cellcolor{black!30} & \cellcolor{black!30} & \cellcolor{black!30} & \cellcolor{black!30} \\
& \greenmark $\rightarrow$\ref{t2:intervention}: \citep{thakkar2023elevating} &   & \cellcolor{black!30} & \cellcolor{black!30} &\cellcolor{black!30}  & \cellcolor{black!30} & \cellcolor{black!30} \\
\midrule

\multirow{4}{*}{\textbf{\deffngprnt}} & \greenmark $\rightarrow$  \ref{t2:intervention}: \citep{szyller2023conflicting}   & \graymark & \graymark & \cellcolor{black!30} & \cellcolor{black!30} & \cellcolor{black!30} & \cellcolor{black!30} \\
&  \redmark $\rightarrow$ \ref{t2:intervention}: \citep{lukas2021deep} &  &  & \cellcolor{black!30} & \cellcolor{black!30} & \cellcolor{black!30}& \cellcolor{black!30} \\
\midrule

\multirow{2}{*}{\textbf{\defdatawm}}  & \redmark $\rightarrow$ \ref{t2:intervention}: \citep{szyller2023conflicting} & \graymark & \graymark & \graymark & \cellcolor{black!30} & \cellcolor{black!30} & \cellcolor{black!30}  \\
\midrule

\multirow{2}{*}{\textbf{\defdiffpriv}} & \greenmark$\rightarrow$ \ref{t1:optimization}: \citep{bu2022practical,wu2023augment,phan2019heterogeneous,phan2020scalable,zhang2022differentially,he2020robustness} & \greenmark$\rightarrow$ \ref{t2:intervention}: \citep{xumitigatingdata,vos2023differentially,ma2019data} & \redmark $\rightarrow$ \ref{t2:intervention}: \citep{szyller2023conflicting} & \greenmark $\rightarrow$ \ref{t2:intervention}: \citep{szyller2023conflicting} & \greenmark $\rightarrow$ \ref{t2:intervention}: \citep{szyller2023conflicting} & \cellcolor{black!30} & \cellcolor{black!30} \\
\midrule

\multirow{4}{*}{\textbf{\deffair}}  & \greenmark: \ref{t1:optimization}: \citep{tran2022fairness,farmur,benz2021robustness,MaNeurips2022,nanda2021fairness,fairOrRobbust,SunFairRobust,li2023wat,lee2024dafa,wei2023cfa} & \greenmark$\rightarrow$\ref{t1:optimization}: \citep{failof,liu2021fairness,shekhar2021fairod,zhang2021towards} & \graymark & \graymark & \graymark & \greenmark$\rightarrow$ \ref{t1:optimization}: \citep{tran2021lagrange,LIU2022108,lowy2023stochastic,dpfair,tran2021differentially,yaghini2024regulation,ding2020differentially,DPLogistic,zhang2021balancing,esipova2023disparate,xu2020removing,tran2023fairdp} & \cellcolor{black!30} \\
& \greenmark$\rightarrow$ \ref{t2:intervention}: \citep{SunFairRobust} &  & &  &  & \greenmark$\rightarrow$ \ref{t2:intervention}: \citep{yaghini2023learning} & \cellcolor{black!30}\\
\midrule

\multirow{2}{*}{\textbf{\defexpl}}  &  \greenmark$\rightarrow$ \ref{t1:optimization}: \citep{lakkaraju2020robust,NEURIPS2019_172ef5a9,Li_2023_CVPR} & \graymark & \graymark & \graymark & \graymark & \greenmark $\rightarrow$ \ref{t1:optimization}: \citep{yang2022differentially}; \ref{t2:intervention}: \citep{DPExplanations} & \graymark \\

\bottomrule

\toprule
\end{tabular}
}
\label{tab:interactions}
\end{table*}


\subsection{Survey of Prior Work}\label{sec:survey}

We describe the defense combinations in the order of appearance along the columns in Table~\ref{tab:interactions}.
%

\noindent\underline{\textbf{\defevasion + \defoutrem.}}
\noindent \defevasion suppresses adversarial examples (as outliers) while \defoutrem suppresses poisons (as outliers) in $\dtrain$. 
Hence, their objectives are aligned. Viewing \defoutrem as out-of-distribution (OOD) generalization, we can modify adversarial training by incorporating noise from the new domain to improve domain generalization~\citep{xin2023connection}. This allows the model to learning robust features for OOD generalization, thereby aligning with \defoutrem (\greenmark: \ref{t1:optimization}).
\citet{hu2023outlier} defend against both poisons and evasion using a bi-level optimization (\greenmark: \ref{t1:optimization}).

\noindent\underline{\textbf{\defevasion + \defmodelwm.}}
Adversarial training, as \defevasionIn, suppresses the influence of backdoors which are used for \defmodelwmPre (\redmark: \ref{t2:intervention})~\citep{szyller2023conflicting}. 
However, generating adversarial-example based watermarks with a higher $\epsilon_{rob}$ than \defevasionIn, can result in an effective combination (\greenmark: \ref{t2:intervention})~\citep{thakkar2023elevating}.

\noindent\underline{\textbf{\defevasion + \defdatawm.}}
\noindent Radioactive data~\citep{sablayrolles2020radioactive}, as \defdatawmPre, adds backdoors as watermarks to $\dtrain$ by perturbing the inputs (similar to adversarial examples). 
Hence, adversarial training will suppress the influence of watermarks used for \defdatawm (\redmark: \ref{t2:intervention})~\citep{szyller2023conflicting}.

\noindent\underline{\textbf{\defevasion + \deffngprnt.}}
Dataset inference~\citep{maini2021dataset} (as \deffngprnt) is effective with \defevasion and incurs an acceptable performance drop (\greenmark: \ref{t2:intervention})~\citep{szyller2023conflicting}. 
We attribute this to the defenses being applied at different stages (in-training vs. post-training), which reduces conflict between them.
On the other hand, a variant of \deffngprnt based on adversarial examples (i.e., ``conferrable examples''), are ineffective when \defevasion is applied for the target or the surrogate model (\redmark: \ref{t2:intervention})~\citep{lukas2021deep}. 
We mark them both separately in Table~\ref{tab:interactions}.

\noindent\underline{\textbf{\defevasion + \defdiffpriv.}}
\citet{hayes2022learning} show that the generalization is worse on combining the objectives of adversarial training (as \defevasionIn) and DPSGD (as \defdiffprivIn), suggesting a conflict (\redmark: \ref{t1:optimization}).

\citet{bu2022practical} modify the minimax objective function of adversarial training to include DPSGD without violating its guarantees (\greenmark: \ref{t1:optimization}).
\citet{wu2023augment} combine randomized smoothing with DPSGD by averaging the gradients of multiple training sample augmentations before clipping to account for the privacy budget of adversarial examples. Both techniques modify the objective function (\greenmark: \ref{t1:optimization}).

Training $\model$ on some public data along with the choice of \defdiffpriv hyperparameters followed by task specific fine-tuning can result in better trade-off (\greenmark: \ref{t1:optimization})~\citep{zhang2022differentially,he2020robustness}.
Other works have considered different optimizations: add \defdiffpriv noise to both input and hidden layers, ensemble adversarial training to add adversarial examples to private $\dtrain$, or modify objective function for \defdiffpriv guarantees on adversarial examples (\greenmark: \ref{t1:optimization})~\citep{phan2020scalable,phan2019heterogeneous}.

\noindent\underline{\textbf{\defevasion + \deffair.}}
\noindent 
Adversarial training (as \defevasionIn) and group fairness (as \deffairIn) have conflicting objectives: \defevasionIn pushes the decision boundary away from $\dtrain$ while \deffairIn brings it closer~\citep{tran2022fairness}.
Also, \defevasionIn increases the disparity among demographic subgroups due to class imbalance in $\dtrain$~\citep{Hu2023RobFair} and long-tailed distribution~\citep{lee2024dafa,benz2021robustness,nanda2021fairness,Hu2023RobFair}. 
Several works modify \defevasionIn's objective function to ensure equitable model behavior across demographic subgroups by assigning higher weight to minority subgroup (\greenmark: \ref{t1:optimization})~\citep{farmur,benz2021robustness,MaNeurips2022,nanda2021fairness,fairOrRobbust,SunFairRobust,lee2024dafa,li2023wat}.
\citet{wei2023cfa} use different training configurations and assigning different weights to different classes to improve class-wise robustness (\greenmark: \ref{t1:optimization}).

\noindent\underline{\textbf{\defevasion + \defexpl.}}
\noindent Adversarial training (as \defevasionIn) improves the interpretability of the gradients~\citep{tsipras2018robustness}.
This suggesting an alignment with explanations~\citep{chalasani20a}.
Also, both defenses can be combined using a minimax objective to construct high fidelity explanations while resisting adversarial examples (\greenmark: \ref{t1:optimization})~\citep{lakkaraju2020robust,NEURIPS2019_172ef5a9,Li_2023_CVPR}.

\noindent\underline{\textbf{\defoutrem + \defdiffpriv.}}
\noindent \defdiffpriv reduces the influence of outliers thereby improving robustness against poisons as shown in several works~\citep{xumitigatingdata,vos2023differentially,ma2019data,JagielskiDPPoison}.
Hence, DPSGD mitigates poisons and the defenses have aligned objective, resulting in an effective combination (\greenmark: \ref{t2:intervention}). 

\noindent\underline{\textbf{\defoutrem + \deffair.}}
\noindent \defoutrem may overly flag data records from the minority groups as outliers for removal, which increases the bias~\citep{shekhar2021fairod}.
This can be corrected by reweighing the scores assigned to outliers to account for sensitive attributes (\greenmark: \ref{t1:optimization})~\citep{failof,liu2021fairness}.
Also, an outlier detector can be trained to minimize the correlation between outlier scores and sensitive attributes (\greenmark: \ref{t1:optimization})~\citep{shekhar2021fairod,zhang2021towards}.

\noindent\underline{\textbf{\defmodelwm + \defdiffpriv.}}
\noindent DPSGD (as \defdiffprivIn) reduces memorization of data records in $\dtrain$ and reduces the impact of backdoors for watermarking (\defmodelwm). Hence, \defmodelwm conflicts with \defdiffpriv (\redmark: \ref{t2:intervention})~\citep{szyller2023conflicting}.

\noindent\underline{\textbf{\defdatawm + \defdiffpriv.}}
\noindent Ideally, DPSGD (as \defdiffprivIn) suppresses watermarks (in \defdatawm), suggesting a conflict. However, empirically, \defdatawm was effective when combined with DPSGD~\citep{szyller2023conflicting}. 
The adversarial example-based watermarks radioactive watermarking~\citep{sablayrolles2020radioactive}, were relatively inliers and not suppressed by DPSGD (\greenmark: \ref{t2:intervention})~\citep{szyller2023conflicting}.

\noindent\underline{\textbf{\deffngprnt + \defdiffpriv.}}
\noindent Szyller and Asokan~\citep{szyller2023conflicting} found that DPSGD (as \defdiffprivIn) and dataset inference (\deffngprnt) do not conflict, though no reason was provided.
We attribute this to defenses being applied in different stages, reducing conflict (\greenmark: \ref{t2:intervention}).

\noindent\underline{\textbf{\defdiffpriv + \deffair.}}
\noindent DPSGD (as \defdiffprivIn) shows disparate behavior over demographic subgroups~\citep{bagdasaryan2019differential}.
Theoretically, it is impossible to design a high utility binary classifier that is private and fair~\citep{cummingsPrivFair,sushantTradeoffs}.
Several works modify the objective function by using fairness constraints, regularization, and game theoretic optimization (\greenmark: \ref{t1:optimization})~\citep{tran2021lagrange,LIU2022108,lowy2023stochastic,tran2021differentially,dpfair,yaghini2024regulation,mozannar2020fair}.
\citet{yaghini2023learning} combine demographic parity regularization with DPSGD, and estimate fairness on a public dataset to avoid consuming extra privacy budget (\greenmark: \ref{t1:optimization}).
Also, functional mechanism adds Laplace noise to the objective function, along with varied noise levels for different subgroups, which reduces discrimination (\greenmark: \ref{t1:optimization})~\citep{ding2020differentially,DPLogistic}.
However, this is limited to the convex objective functions (e.g., logistic regression).
\citet{esipova2023disparate} attribute unfairness in DPSGD to the differences in unclipped and clipped gradient directions.
Subsequently, several works have used proposed variable gradient clipping to minimize discriminatory behavior while maintaining utility (\greenmark: \ref{t1:optimization})~\citep{xu2020removing,tran2023fairdp,zhang2021balancing}.
\citet{yaghini2023learning} use PATE framework to apply fairness constraints and \defdiffpriv noise in the aggregated votes from the teacher's ensemble. Both fairness and privacy are applied in pre-training (\greenmark: \ref{t2:intervention}).

\noindent\underline{\textbf{\defdiffpriv + \defexpl.}}
\noindent The objectives of these defenses are inherently conflicting: \defdiffpriv hides information to minimize leakage while \defexpl releases additional information to improve comprehensibility~\citep{balancingRights}.
\citet{yang2022differentially} train an autoencoder with \defdiffpriv (functional mechanism). This autoencoder is used to generate data records and compute counterfactuals that satisfy \defdiffpriv via the post-processing property (\greenmark: \ref{t1:optimization}). 
\citet{DPExplanations} propose an adaptive DPSGD algorithm to generate high-quality explanations without consuming $\epsilon_{dp}$, by reusing past explanations for similar data records (\greenmark: \ref{t2:intervention}).

\section{Insights from Systematization}\label{sec:application}

We identify unexplored combinations (\Snospace\ref{sec:unexplored}), requirements for an ideal technique, limitations of prior techniques (\Snospace\ref{sec:desiderata}), and design a new technique (\Snospace\ref{sec:approach}).

\subsection{Unexplored Combinations}\label{sec:unexplored}

We identify 14 unexplored combinations (\graymark\xspace in Table~\ref{tab:interactions}):
\begin{enumerate*}[itemjoin={;\xspace},label={(\roman*)}]
\item \text{\defoutrem} with \{\text{\defmodelwm}, \text{\defdatawm}, \text{\deffngprnt}, \text{\defexpl}\} 
\item \text{\defmodelwm} with \{\text{\defdatawm}, \text{\deffngprnt}, \text{\deffair}, \text{\defexpl}\}
\item \text{\defdatawm} with \{\deffngprnt,\deffair,\text{\defexpl}\}
\item \text{\deffngprnt} with \{\text{\deffair}, \text{\defexpl}\}
\item \text{\deffair} with \text{\defexpl}.
\end{enumerate*}

\begin{takeaway}
\textbf{Takeaway}: Unexplored combinations reveal research gaps and opportunities for effective technique design.
\end{takeaway}
\noindent We revisit these combinations in our evaluation (\Snospace\ref{sec:setup} and~\Snospace\ref{sec:evaluation}).

\subsection{Evaluating Combination Techniques}\label{sec:desiderata}

Our systematization reveals that some techniques may lead to ineffective combinations.
We outline requirements for an ideal technique, and identify limitations in prior work.

\noindent\textbf{\underline{Requirements.}} 
A combination technique should alow practitioner to quickly determine whether a combination can be effective combined.
Empirical evaluation to determine the effectiveness of a combination, while definitive, can be expensive, especially when multiple defenses are involved.
An ideal combination technique should be:
\begin{enumerate*}[label=\textbf{R\arabic*}, itemjoin={;\xspace}]
\item\label{n1:effective}\textbf{(Accurate)} correctly identifies whether a combination is effective or not
\item\label{n2:scalable}\textbf{(Scalable)} allows two or more to be combined simultaneously
\item\label{n3:noninvasive}\textbf{(Non-invasive)} does not require modifying defenses, easing adoption and removing the need for expert knowledge
\item\label{n4:general}\textbf{(General)} applicable to various defenses.
\end{enumerate*}

\begin{takeaway}
\textbf{Takeaway:} Combination techniques, including newly proposed ones, should be evaluated for \ref{n1:effective}-\ref{n4:general}.
\end{takeaway}

\noindent\textbf{\underline{Limitations of Prior Techniques.}} We summarize the limitations of existing techniques (\ref{t1:optimization}-\ref{t2:intervention}) as per the requirements (\ref{n1:effective}-\ref{n4:general}) in Table~\ref{tab:requirements}. We use \xmark for requirement not satisfied, \smark for partially satisfied, and \cmark for fully satisfied.

\setlength\tabcolsep{4pt}
\begin{wraptable}[8]{r}{0.5\textwidth}
\centering
\vspace{-0.75cm}
\caption{Requirements satisfied by various techniques: \xmark $\rightarrow$ Not satisfied; \smark $\rightarrow$ Partially satisfied; \cmark $\rightarrow$ fully satisfied.}
\footnotesize
\resizebox{0.5\textwidth}{!}{ 
\begin{tabular}{c|c|c|c|c}
\bottomrule
    
\toprule
 \textbf{Technique} & \ref{n1:effective}  & \ref{n2:scalable}  & \ref{n3:noninvasive}  & \ref{n4:general}\\ 
 & (Accurate) & (Scalable) &(Non-Invasive) & (General)\\
\bottomrule
    
\toprule
\ref{t1:optimization} & \smark & \xmark & \xmark & \xmark \\  
\ref{t2:intervention} & \smark & \cmark & \cmark & \cmark\\
\bottomrule
    
\toprule
\end{tabular}
}
\label{tab:requirements}
\end{wraptable}

\noindent\textbf{\ref{t1:optimization} (Optimization)} where modifying the objective function for training, followed by hyperparameter tuning, can result in an effective combination.
However, this often results in a trade-off between effectiveness of constituent defenses and model utility. 
Hence, we mark \ref{t1:optimization} as partially accurate (\ref{n1:effective} $\rightarrow$ \smark).
This trade-off also explains why prior works have struggled to scale beyond two defenses (\ref{n2:scalable} $\rightarrow$ \xmark).
Furthermore, some defenses are not applicable for \ref{t1:optimization} (e.g., \defexpl and \deffngprnt), and  require modifications or non-standard variants (\ref{n3:noninvasive} $\rightarrow$ \xmark).
Finally, these optimizations are tailored to specific defenses being combined, and do not apply to other defenses. 
They are also specific to some models (e.g., logistic regression with \defdiffpriv), and do not translate to other models (e.g., neural networks).
Hence, \ref{t1:optimization} has limited applicability (\ref{n4:general} $\rightarrow$ \xmark).

\noindent\textbf{\ref{t2:intervention} (Mutually Exclusive  Placement)} can apply up to one defense in each of the three stages, thus, making it scalable (\ref{n2:scalable} $\rightarrow$ \cmark).
Defenses do not need any modification (\ref{n3:noninvasive} $\rightarrow$ \cmark), and the combination technique is applicable to all types of defenses (\ref{n4:general} $\rightarrow$ \cmark).
However, the combinations may not be effective: (i) a defense in a later stage of the pipeline can conflict with earlier ones (false negatives)~\citep{szyller2023conflicting}, and (ii) it rules out defenses in the same stage that do not conflict (false positives see \Snospace\ref{sec:evaluation}). Hence, this may incorrectly identify effective combinations (partially accurate \ref{n1:effective} $\rightarrow$ \smark).

\begin{takeaway}
\textbf{Takeaway.} \emph{Neither technique satisfies all the requirements.}
\ref{t2:intervention} is promising as it satisfies \ref{n2:scalable}, \ref{n3:noninvasive}, and \ref{n4:general}, but not \ref{n1:effective} (incurs false positives and false negatives).
\end{takeaway}

\subsection{\method: Design}\label{sec:approach}

From our systematization, we identify that \ref{t2:intervention} overlooks underlying causes for conflicts, leading to false positives/negatives.
\emph{Can we address the limitations of \ref{t2:intervention} and improve \ref{n1:effective}?}
Recall from \Snospace\ref{sec:framework} that conflicts arise when (i) an early-stage defense uses a risk which is mitigated by a later-stage defense, or (ii) changes by an early-stage defense is overridden by a later-stage defense.
We conjecture that by accounting for these underlying causes, we can meet \ref{n1:effective}. 
We present \method, a principled technique to identify effective defense combinations, by accounting for the reasons underlying conflicts
\tikzstyle{red} = [rectangle, rounded corners, minimum width=1.5cm, minimum height=0.75cm,text centered, draw=black]
\tikzstyle{green} = [rectangle, rounded corners, minimum width=1.5cm, minimum height=0.75cm,text centered, draw=black]
\tikzstyle{process} = [rectangle, minimum width=1.5cm, minimum height=0.75cm, text centered, draw=black, fill=gray!15]

\begin{wrapfigure}[15]{r}{0.5\textwidth}
\vspace{-0.5cm}
\centering
\resizebox{0.4\textwidth}{!}{
\begin{tikzpicture}[node distance=1cm]
\node (stage) [process, align=center] {\footnotesize \ref{step1}: Same Stage?};

\node (global) [process, below of=stage, xshift=-1.8cm, yshift=-0.5cm, align=center] {\footnotesize \ref{step2}: \defensetwo $\rightarrow$ local/no change?};

\node (relies) [process, below of=stage, xshift=2cm, yshift=-0.5cm] {\footnotesize \ref{step3}: \defenseone uses \risk?};

\node (red4) [red, below of=global, yshift=-0.5cm, xshift=-0.75cm] {\footnotesize \redmarkOurs};
\node (green3) [green, right of=red4, xshift=0.6cm] {\footnotesize \greenmarkOurs};

\draw [ultra thick, ->] (global) -- node[anchor=east] {no} (red4);
\draw [ultra thick, ->] (global) -- node[anchor=west] {yes} (green3);

\draw [ultra thick, ->] (stage) -- node[anchor=east] {yes} (global);
\draw [ultra thick, ->] (stage) -- node[anchor=west] {no} (relies);

\node (s11) [process, left of=relies, xshift=0.2cm, yshift=-1.5cm, align=center] {\footnotesize \textbf{\ref{step4}}: \defensetwo protects\\\footnotesize against \risk?};

\node (red1) [red, below of=s11, yshift=-0.5cm, xshift=-0.75cm] {\footnotesize \redmarkOurs};

\draw [ultra thick, ->] (relies) -- node[anchor=east] {yes} (s11);
\draw [ultra thick, ->] (s11) -- node[anchor=east] {yes} (red1);

\node (green1) [green, right of=red1, xshift=1cm] {\footnotesize \greenmarkOurs};

\draw [ultra thick, ->] (s11) -- node[anchor=west] {no} (green1);
\draw [ultra thick, ->] ([xshift=0.75cm]relies.south) -- node[anchor=west] {no} ([xshift=0.35cm]green1.north);





\end{tikzpicture}
}
\caption{Flowchart depicting various steps in \method to identify conflict between \defenseone and \defensetwo (\defensetwo is applied after \defenseone).}
\label{fig:flowchart}
\end{wrapfigure}
\noindent\textbf{\underline{Methodology to Derive \method.}} We start with the na\"ive technique and modify it to include the underlying causes for conflicts among defenses.
We iterate over the design of \method using prior work from \Snospace\ref{sec:systematization}, and evaluate the final design on unexplored combinations (see \Snospace\ref{sec:evaluation}).

\noindent\textbf{\underline{\method Description.}} We describe \method using the example of combining two defenses, \defenseone and \defensetwo which protect against \riskone and \risktwo respectively, and later discuss how to extend to more than two defenses.
Following prior work~\citep{duddu2023sok}, we refer to unintended interactions between a defense and a risk if the defense either increases or decreases the susceptibility to an unrelated risk (e.g., \defenseone and \risktwo).

We start by identifying variants of each of the defenses across pre-, in-, and post-training stages (see Table~\ref{tab:defenses}).
We compare each variant of \defenseone with that of \defensetwo, and use \greenmarkOurs\xspace for alignment, and \redmarkOurs\xspace for conflict.
Assuming \defenseone is applied first and then \defensetwo, we follow the steps below \emph{in sequence}:
\begin{enumerate}[label=\textbf{S\arabic*}, leftmargin=*, wide,labelindent=0pt]
\item\label{step1} Are \defenseone and \defensetwo applied in the same stage?\\
\textbullet\xspace If yes, go to \ref{step2} \textbullet\xspace Else, go to \ref{step3}

\item\label{step2} The type of changes made by the defenses determines whether there is a conflict. We classify the changes as \emph{global}, \emph{local}, and \emph{none}.
\emph{Global changes} modify $\model$ (e.g., training with a regularization term, pruning) or transform all records in $\dtrain$ (e.g., synthetic data generation for DP or fairness during pre-training).
\emph{Local changes} affect specific data records (e.g., adding watermarks in pre-training or modifying certain predictions in post-training).
\deffngprntPost and \defexplPost make \emph{no changes} to $\model$ and $\dtrain$.
\begin{itemize}[leftmargin=*]
    \item If \defenseone makes global/local/no changes while \defensetwo makes local/no changes, we mark this as \greenmarkOurs\xspace.\\  
    \noindent\textbf{Rationale:} \emph{Changes by \defenseone will not interfere with local/no changes by \defensetwo, as \defenseone is applied first. Hence, no conflict.} 

    \item If \defenseone makes global/local/no changes while \defensetwo makes global changes, mark as \redmarkOurs\xspace.\\
    \noindent\textbf{Rationale:} \emph{Global changes by \defensetwo will override changes by \defenseone, thereby reducing its effectiveness. This is called catastrophic forgetting when the defenses are applied sequentially during training~\citep{kemker2018measuring,szyller2023conflicting}. This is a conflict.} 
\end{itemize}

\item\label{step3} \defenseone and \defensetwo are in different stages. Does \defenseone use a risk \risk as part of the defense (e.g., watermarking uses backdoors)? 
\begin{itemize}[leftmargin=*]
\item If yes, go to \ref{step4}.
\item If no, mark as \greenmarkOurs.\\
\noindent \textbf{Rationale:} \emph{If \defenseone does not use \risk, the susceptibility to \risk will not be impacted after applying \defensetwo. Hence, \defenseone and \defensetwo are unlikely to interfere with each other.}

\end{itemize}

\item\label{step4} Does \defensetwo protect against \risk either explicitly or via unintended interaction?
\begin{itemize}[leftmargin=*]
    \item If yes, mark as \redmarkOurs.\\
    \noindent \textbf{Rationale:} \emph{Since \defenseone uses \risk (either explicitly or via unintended interaction), \defensetwo will reduce susceptibility to \risk making \defenseone less effective. Hence, there is a conflict.}
    \item If no, mark as \greenmarkOurs.\\
    \noindent \textbf{Rationale:} \emph{\defenseone and \defensetwo are unlikely to interfere with each other. Hence, there is no conflict.}
\end{itemize}
\end{enumerate}
We summarize the steps in \method in Figure~\ref{fig:flowchart}.
\method evaluates combination effectiveness based solely on the effectiveness of the constituent defenses, without considering the model utility. We revisit model utility in \Snospace\ref{sec:discussions}. \change{We also present a formal analysis of \method covering consistency, soundness, and completeness in Appendix~\ref{app:formal}.}

\noindent\textbf{\underline{Note on \defdiffpriv.}} 
Combining \defdiffpriv with other defenses does not consume additional privacy budget.
Any modification to $\dtrain$ (e.g., adding watermarks) is done within the privacy boundary, and does not consume privacy budget.
Any defense applied after \defdiffpriv is ``free'' (\defdiffpriv's post-processing property).

\noindent\textbf{\underline{Extending Beyond Two Defenses (Multi-way Combinations).}}
\change{To extend \method beyond two defenses, our algorithm decomposes the multi-way combination into pairwise comparisons, invoking the original \method flowchart (Figure~\ref{fig:flowchart}) to check for conflicts.
Determining conflict for multi-way combinations from pairwise combinations is principled and not limiting. We have to consider two possible cases when decomposing multi-way combinations into pairwise combinations:
\begin{itemize}[leftmargin=*]
    \item \textbf{Case 1: Defenses in different stages.} When we combine three defenses in different stages, checking conflicts among all pairwise combinations is fine since the order in which we apply defenses is fixed. Any conflict detected among any pair, will be marked as an overall conflict (marked as \redmarkOurs).
    \item \textbf{Case 2: More than one defense in one stage.} We have to check all possible permutations of the defenses in a given stage, and determine whether there is a conflict. If there is no conflict, we combine with other stage defenses, and check for the conflicts.
\end{itemize}

We now present our algorithm to evaluate conflict for multi-way combinations, and assume that we have a set of defenses—partitioned into three ordered stages. 
We iterate through each stage in order, and skip stages with no defenses:
\begin{enumerate}[label={\textbf{M\arabic*}},leftmargin=*]
\item \textbf{If stage has single defense:} For the stage $s$, we consider the defense as $\defense_s$ and check its compatibility with defenses in other stages. $\defense_s$ is \defenseone, \defensetwo or \defensethree depending on $s$. 
\item \textbf{If stage has multiple defenses:} For each stage with multiple defenses:
\begin{enumerate}[label={\textbf{M2.\arabic*}}]
\item \textbf{Permutation Checking:} Consider all permutations of the defenses. For each permutation, check every consecutive pair of defenses using \method flowchart to detect conflicts for two defenses.
\item \textbf{Pruning and Selection}: If a pair in a permutation causes a conflict, discard that permutation and prune others containing that conflicting pair (e.g., for defenses: A, B, and C, if  ``AB'' conflict, then no need to check for ``ABC'' or ``CAB''). If a conflict-free permutation is found, treat the entire sequence as a single composite defense, $\defense_s$ (e.g., let us say ``CAB'' does not conflict).
\end{enumerate}
\item \textbf{Sequential Composition}: Treat the resulting composite defense as an atomic unit (e.g., \defenseone = ``CAB'') and check for conflicts with the next stage's resolved defenses (e.g., \defensetwo = ``DE''), using \method flowchart between \defenseone and \defensetwo. In other words, we check whether any of the constituent defenses in \defenseone use a risk which is mitigated by constituent defenses in \defensetwo (\ref{step3} in Figure~\ref{fig:flowchart}).
\item \textbf{Termination:} If any invocation of \method for pairwise check indicates a conflict, terminate and report as a conflict (marked as \redmarkOurs). Else, indicate as alignment (marked as \greenmarkOurs).
\end{enumerate}
}

\change{We present a formal analysis of the algorithm to extend \method for more than two defenses in Appendix~\ref{app:formalMultiway}.}
Having discussed the \method's design, we comprehensively evaluate \method across \ref{n1:effective}-\ref{n4:general}.
\section{Experimental Setup}\label{sec:setup}


\subsection{Datasets and Models}\label{sec:dataset}

We use two image datasets: \texttt{FMNIST} and \texttt{UTKFACE}.
\texttt{FMNIST} consists of 28x28 grayscale images of ten clothing types, with 60,000 training and 10,000 testing images. We classify these using a two layer CNN with 16 and 32 filters, ReLU activation, and a fully connected layer for ten-class classification.
\texttt{UTKFACE} includes 48x48 RGB images, classifying individuals as young (under 30), with 11,852 training and 10,667 testing images. It also includes the sex of the individuals as a sensitive attribute. 
We use a VGG16 model with a fully connected layer for binary classification.

We choose \texttt{FMNIST} since all of the defenses we consider for evaluation (\Snospace\ref{sec:chooseDef}) have used it for evaluation.
We selected \texttt{UTKFACE} because many defenses effective for \texttt{FMNIST} are likely to be applicable to it as well, given that both are image datasets. Also, \texttt{UTKFACE} includes sensitive attributes, making it suitable for \deffair.

\subsection{Revisiting Defenses}\label{sec:revistdefenses}

Since the na\"ive technique and \method apply defenses at different ML pipeline stages, we revisit and categorize defenses in \Snospace\ref{sec:defenses} by the stage they are applied in.
For each defense from \Snospace\ref{sec:defenses}, we specify the variants in pre-training (``<defense>.\textbf{\texttt{Pre}}''), in-training (``<defense>.\textbf{\texttt{In}}''), and post-training (``<defense>.\textbf{\texttt{Post}}''). 
For additional context, we indicate the impact of applying a defense on \phiacc compared to a ``no defense'' baseline, where ``\downmark'' is a decrease, ``\samemark'' is no effect, and ``\upmark'' is an increase.

\noindent\textbf{\underline{Evasion robustness (\defevasion})}
\begin{itemize}[leftmargin=*]
\item \textbf{\defevasionPre (Data Augmentation)} where adding transformations of training data records improves robustness~\citep{cutmix,zhang2018mixup,devries2017cutout,rebuffi2021data} (but see \Snospace\ref{sec:chooseDef}). 
This improves \phiacc (\upmark) by acting as regularization~\citep{cutmix,zhang2018mixup,devries2017cutout,rebuffi2021data}.

\item \textbf{\defevasionIn (Adversarial Training)} modifies the objective function to minimize the maximum loss from adversarial examples~\citep{madry2018towards,trades}:
$L_{advtr} = \min_{\theta} \frac{1}{|\dtrain|} \sum_{x,y \in \dtrain} \max_{\|\delta\| \leq \epsilon_{rob}}\ell(\model(x+\delta), y)$.
Alternatively, randomized smoothing modifies the training and inference to obtain certified robustness of $\model$~\citep{cohen19c,lecuyer2019certified}.
These defenses decrease \phiacc (\downmark)~\citep{trades,tsipras2018robustness}.

\item \textbf{\defevasionPost (Input Processing)} removes adversarial perturbations before passing them to $\model$ (e.g., using generative models~\citep{nie2022DiffPure,song2018pixeldefend} or input encoding~\citep{buckman2018thermometer,guo2018countering,Das2017KeepingTB}) or checks for adversarial examples using statistical tests~\citep{grosse2017statistical}.
Defenses which modify input images using generative models decrease \phiacc (\downmark)~\citep{nie2022DiffPure,song2018pixeldefend,guo2018countering,Das2017KeepingTB}. If the input transformation is small, the decrease in \phiacc is negligible (\samemark)~\citep{buckman2018thermometer,grosse2017statistical}.
\end{itemize}

\noindent\textbf{\underline{Poison robustness (\defoutrem})}
\begin{itemize}[leftmargin=*]
\item \textbf{\defoutremPre (Data Sanitization)} includes detecting and removing outliers in $\dtrain$ (e.g., using Shapley values~\citep{jia2021scalability,jia2019efficient,doan2020februus} or anomaly detection~\citep{cretu2008,paudice2018detection,spectralsig,roni,chen2018detecting}), followed by retraining.
As the outliers are memorized and contribute to \phiacc, their removal degrades \phiacc (\downmark)~\citep{jia2021scalability,jia2019efficient}.

\item \textbf{\defoutremIn (Fine-tuning)} updates $\model$ to minimize outlier influence. 
This includes distillation to reduce the influence of poisons~\citep{distillationDefense} or fine-tuning on a poison-free dataset~\citep{diakonikolas19a,zhu2023neural,xu2019l_dmi,liu2020peer,patrini2017making}.
These do not impact \phiacc (\samemark).

\item \textbf{\defoutremPost (Pruning)} reduces the effectiveness of backdoors by removing some model parameters based on the observation that poisoned and clean samples have different activations~\citep{liu2018fine,wu2021adversarialBackdoor,zheng2022preactivation,zheng2022data,li2023reconstructive}.
This degrades \phiacc (\downmark).
\end{itemize}

\noindent\textbf{\underline{Model Watermarking (\defmodelwm})}
\begin{itemize}[leftmargin=*]
\item \textbf{\defmodelwmPre (Backdoors)} uses backdoor watermarks in $\dtrain$~\citep{adiPaper,zhangWM,jiaEntabgled,uchida}. 
These are designed to maintain \phiacc (\samemark).

\item \textbf{\defmodelwmIn (Optimization)} updates the original objective function to include watermark behavior~\citep{bansal2022certified,bagdasaryan2021blind}. 
For instance, certified watermarking adds Gaussian noise to watermarks (added to $\dtrain$) to get certification on watermark accuracy~\citep{bansal2022certified}.
Also, backdoors can be introduced through regularization, which can be repurposed for watermarking~\citep{bagdasaryan2021blind}.
This degrades \phiacc (\downmark).

\item \textbf{\defmodelwmPost (API)} modifies predictions to embed watermarks~\citep{dawn} which are used by \adv to train the surrogate model.
These are designed to maintain \phiacc (\samemark).
\end{itemize}

\noindent\textbf{\underline{Fingerprinting (\deffngprnt}).} All fingerprints are post-training schemes (denoted as \textbf{\deffngprntPost}).
No retraining or modification of $\model$ is required and hence, \deffngprnt has no effect on \phiacc (\samemark).

\noindent\textbf{\underline{Data watermarking (\defdatawm}).} All the current schemes are during pre-training (\textbf{\defdatawmPre}), and are designed to maintain \phiacc (\samemark).
The difference between \defmodelwm and \defdatawm is how a model trained from scratch on $\dtrain$ is classified: \defdatawm flags it for unauthorized data use while \defmodelwm classifies it as independently trained.

\noindent\textbf{\underline{Differential privacy (\defdiffpriv})}
\begin{itemize}[leftmargin=*]
\item \textbf{\defdiffpriv.\textbf{\texttt{Pre}} (Private Data)} from generative models with \defdiffpriv constraints, that can be used for downstream tasks instead of $\dtrain$~\citep{hu2023sok,Xie2018DifferentiallyPG,torkzadehmahani2019dp,zheng2023differentially}. 
This decreases \phiacc (\downmark).
\item \textbf{\defdiffpriv.\textbf{\texttt{In}} (DPSGD)} trains $\model$ by adding carefully computed noise to the gradients to minimize the influence of individual data records on $\model$~\citep{abadi2016deep}.
Private aggregation of teacher's ensembles (PATE)~\citep{papernot2017semisupervised} is another framework for DP where multiple teacher models are trained on disjoint private datasets, while a student model is trained on a public dataset with labels annotated via noisy voting from the teacher models.
These defenses decrease \phiacc (\downmark)~\citep{dpacc}.
\item \textbf{\defdiffprivPost (Output Perturbation)} includes adding calibrated noise to the output of empirical risk minimization objective~\citep{chaudhuri2011differentially}.
This decreases \phiacc (\downmark).
The theoretical guarantees are poorer than other DP defenses and \defdiffprivPost requires the objective function to be convex.
We omit this since it does not cover neural networks.
\end{itemize}

\noindent\textbf{\underline{Group fairness (\deffair})} 
\begin{itemize}[leftmargin=*]
\item \textbf{\deffairPre (Fair Data)} modifies $\dtrain$ to reduce bias in the downstream model~\citep{datapreproc,optpreproc,lfr,feldman2015certifying}. 
This degrades \phiacc (\downmark).

\item \textbf{\deffairIn (Regularization)} penalizes violating fairness constraints~\citep{agarwal2018reductions,agarwal2019fair,celis2019classification,prejremover}.
This degrades \phiacc (\downmark)~\citep{zhang2018mitigating,louppe2017learning,fairVacc1}.

\item \textbf{\deffairPost (Calibration)} adjusts the threshold over the predictions to ensure that the prediction probabilities accurately reflect the true likelihood across each demographic group~\citep{pleiss2017fairness,hardt2016equality,discClassification,geyik2019fairness,salvador2022faircal,kull17a,johnson18a}
This degrades \phiacc (\downmark)~\citep{pleiss2017fairness}.
\end{itemize}

\noindent\textbf{\underline{Explanations (\defexpl}).} \defexpl are post-training defenses (\textbf{\defexplPost}) which does not require retraining, and hence does not degrade \phiacc (\samemark).

\noindent We summarize the defenses in Appendix~\ref{app:notations}: Table~\ref{tab:defenses}.

\subsection{Choosing Defenses for Evaluation}\label{sec:chooseDef} 

To select defenses for our evaluation, we began with those in \Snospace\ref{sec:revistdefenses} (summarized in Appendix~\ref{app:notations}: Table~\ref{tab:defenses}).
We remove defenses which are not robust: \defevasionPost (Input Processing) and \defoutremPre (Data Sanitization)~\citep{kang2024diffattack,koh2022stronger}. 
We then evaluate the remaining defenses and exclude those which were ineffective on our datasets: \defevasionPre (Data Augmentation)~\citep{cutmix,zhang2018mixup,devries2017cutout}, \defdiffprivPre (Private Data)~\citep{zheng2023differentially}, \deffairPre (Fair Data)~\citep{lfr}, and \deffairPost (Calibration)~\citep{pleiss2017fairness}.
\defdiffprivPre, \deffairPre, and \deffairPost, were designed for tabular datasets but ineffective on our image datasets.
We speculate about these defenses in \Snospace\ref{sec:discussions}.

We are left with eleven defenses:
\begin{enumerate*}[label=(\roman*),itemjoin={,\xspace}]
\item \defevasionIn (adversarial training)
\item \defoutremIn (fine-tuning)
\item \defoutremPost (model pruning)
\item \defmodelwmPre (backdoor watermarks)
\item \defmodelwmIn (watermarks via objective function)
\item \defmodelwmPost (API-based watermarks)
\item \defdatawmPre (backdoor watermarks)
\item \deffngprntPost (dataset inference)
\item \defdiffprivIn (DPSGD)
\item \deffairIn (regularization)
\item \defexplPost (attribution).
\end{enumerate*}
We get 55 pairwise combinations from them but we remove combinations among defenses with the same objective: three combinations among watermarking (\defmodelwmPre, \defmodelwmIn, \defmodelwmPost), three for \deffngprntPost with \defmodelwmPre, \defmodelwmIn, \defmodelwmPost, and one for \defoutremIn and \defoutremPost. This leaves us with 48 combinations.

\subsection{Metrics and Implementations}\label{sec:metrics}

We describe the metrics for evaluating the effectiveness of each defense, and the implementations taken from publicly available code from prior work. We measure \phiacc on $\dtest$ for all defenses.
Our implementations for defenses are based on the state-of-the-art (\deffngprnt, \defoutrem), standard libraries (\defdiffpriv, \deffair, \defexpl), or seminal work (\defevasion, \defdatawm, \defmodelwm).
We use the standard hyperparameters which are either from the literature or the library documentation, such that the resulting individual defenses are effective (Table~\ref{tab:tab_eval_standalone}).
When combining defenses, we use the same hyperparameters, but revisit hyperparameter tuning for defenses in combination (see \Snospace\ref{sec:hyperparam}).
We report the mean and standard deviation across five runs. 

\noindent\textbf{\underline{Evasion Robustness (\defevasionIn).}}
We use the accuracy on $\mathcal{D}_{rob}$ which is obtained by replacing data records in $\dtest$ with the adversarial variants: 
\begin{equation*}
\begin{split}
\phi_{\text{robacc}}(\model_{\text{\defevasion}}, \mathcal{D}_{\text{rob}}) 
= \frac{1}{|\mathcal{D}_{\text{rob}}|}
\sum_{(x, y) \in \mathcal{D}_{\text{rob}}} 
\mathbb{I}\left\{ \hat{\model}_{\text{\defevasion}}(x) = y \right\}
\end{split}
\end{equation*}
Ideally, $\phi^{\text{\defevasion}}_{\text{robacc}}$ should be close to \phiacc.
We use the original implementation of TRADES~\citep{trades} and an implementation of AutoAttack~\citep{autoattack} in SecML library~\citep{secml} from \citet{attackFailures}. 
As we evaluate effectiveness of \defevasionIn using attacks, poorly optimized attacks can falsely suggest defense effectiveness~\citep{carlini2019evaluating,carlini2017adversarial,tramer2020adaptive}.
We individually optimize these attacks for evaluation with defenses and their combinations, following \citet{attackFailures} to address failures identified by various indicators (e.g., poor optimization).
We evaluate across various AutoAttack variants by (a) modifying the loss function: cross entropy (CE) and difference of logits ratio (DLR), (b) applying expectation over transformations (EoT), and (c) using random starts. For \texttt{FMNIST}, we use DLR, CE, DLR+EoT, and CE+EoT. For \texttt{UTKFACE}, we use CE and CE+EoT, as DLR is not applicable for binary classification.
We report the best attack (least $\phi_{robacc}$).


\noindent\textbf{\underline{Outlier Removal (\defoutrem).}}
We compute accuracy on $\mathcal{D}_{bd}$ (from adding backdoors to records in $\dtest$):
\begin{equation*}
\begin{split}
\phi_{\text{ASR}}(\model_{\text{\defoutrem}}, \mathcal{D}_{\text{bd}}) 
= \frac{1}{|\mathcal{D}_{\text{bd}}|}
\sum_{(x, y) \in \mathcal{D}_{\text{bd}}} 
\mathbb{I}\left\{ \hat{\model}_{\text{\defoutrem}}(x) = y \right\}
\end{split}
\end{equation*}
where $y_t$ is the target label chosen by \adv. Ideally, $\phi_{\text{ASR}}$ should be zero. 
We use BadNets~\citep{gu2017badnets} to generate poisons by adding a white patch of size 5x5 to the images, applied to 10\% of $\dtrain$. 
For \defoutremIn (Fine-tuning), we fine-tune the last layers of $\model$ using random sample of 10\% of $\dtrain$ without poisons~\citep{sha2022fine}. 
For \defoutremPost (Pruning), we use the implementation from \citet{zheng2022preactivation}. 
We sweep pruning thresholds from 0.6 to 1.5, in increments of 0.05, to get a model with highest \phiacc and lowest $\phi_{ASR}$.

\noindent\textbf{\underline{Model-Watermarking (\defmodelwm).}}
We use the accuracy on $\mathcal{D}_{\text{wmM}}$ which is obtained by adding watermarks to data records in $\dtest$. We compute this \emph{watermark accuracy} as: 
\begin{equation*}
\begin{split}
\phi_{\text{wmacc}}(\model_{\text{\defmodelwm}}, \mathcal{D}_{\text{wmM}}) 
= \frac{1}{|\mathcal{D}_{\text{wmM}}|}
\sum_{(x, y) \in \mathcal{D}_{\text{wmM}}} 
\mathbb{I}\left\{ \hat{\model}_{\text{\defmodelwm}}(x) = y \right\}
\end{split}
\end{equation*}
where $y_m$ represents the target labels for watermarked records. 
Ideally, $\phi_{\text{wmacc}}$ should be 100\% if the model is successfully watermarked. 
For \defmodelwmPre (Backdoor), we use BadNets~\citep{gu2017badnets}, similar to Szyller and Asokan~\citep{szyller2023conflicting}, by adding a white patch of size 5x5 to 10\% of the images in $\dtrain$.
For \defmodelwmIn (Modifying Loss), we use the certified neural network watermarking implementation by \citet{bansal2022certified}. 
For \defmodelwmPost (API), we use DAWN~\citep{dawn}, which flips a fraction of the predictions from target model as watermarks, which is later used to train the surrogate model. Following the original work~\citep{dawn}, we apply the watermark to 0.2\% of the predictions.
Unlike other watermarking schemes, we compute $\phi_{\text{wmacc}}$ on the surrogate model and not the target model.

\noindent\textbf{\underline{Fingerprinting (\deffngprntPost).}}
We use dataset inference~\citep{maini2021dataset} as our fingerprinting scheme which extracts feature embeddings from $\model$, and trains a classifier to distinguish between $\dtrain$ and $\dtest$. A model is considered stolen if the distance of its embeddings is similar to $\model$ with high confidence, and verification is successful if the p-value < 0.05.
We use $\phi_{\text{pval}}$ as the metric following Szyller and Asokan~\citep{szyller2023conflicting}.
We use the step size of 1.0 for $L_{1}$ attack, 0.01 for $L_{2}$ attack, and 0.001 for $L_{inf}$, and 50 samples for computing p-value from the confidence regressor model.

\noindent\textbf{\underline{Data-Watermarking (\defdatawmPre).}} To determine if a dataset was used to train a model, we compare the posterior probability of 100 watermarked testing samples against 100 benign ones using a pairwise t-test~\citep{li2020open}. We then calculate the rate of successful detection ($\phi_{\text{rsd}}$), which reflects the percentage of correctly identified watermarked samples from $\mathcal{D}_{\text{wmD}}$ ($\dtest$ with watermarks). 
Watermarks are generated using BadNets~\citep{gu2017badnets} where 10\% of $\dtrain$ is watermarked, and we use verification code from \citet{li2020open} to compute $\phi_{\text{rsd}}$. 
Ideally, $\phi_{\text{rsd}}$ should be 100\% for watermarked models.

\noindent\textbf{\underline{Differential Privacy (\defdiffprivIn).}}
We use $\phi_{\text{dp}} = {\epsilon_{dp}}$, following Szyller and Asokan~\citep{szyller2023conflicting}, where ideally, we want a low $\epsilon_{dp}$. We use the implementation from Opacus library~\citep{yousefpour2021opacus} with a noise multiplier of 1.0 and gradient norm clipping of 1.0 as used in their tutorial for \texttt{MNIST}.

\noindent\textbf{\underline{Group Fairness (\deffairIn).}}
We measure fairness using the \emph{equalized odds gap} on $\dtest$ for sensitive attributes $S$ and model predictions $\hat{Y}$, given by: $\phi^{\text{\deffair}}_{\text{eqodd}} = P(\hat{Y}=\hat{y} | S=0,Y=y) - P(\hat{Y}=\hat{y} | S=1,Y=y)$
\noindent$\forall (\hat{y},y)\in\{0,1\}^2$ where an ideal value of zero indicates perfect fairness.
For \deffairIn (Regularization), we use code from the fair fairness benchmark that adds a regularization term to penalize equalized odds violations~\citep{han2023ffb}.
We set the regularization hyperparameter $\lambda=1$ which was sufficient to reduce $\phi_{eqodds}$ with $\sim 2\%$ drop in \phiacc.

\noindent\textbf{\underline{Explanations (\defexplPost).}}
We assess the quality fo explanations using \emph{convergence delta}, measures the error between the explanation for a data records and a baseline~\citep{kokhlikyan2020captum}. We report the average convergence delta across all $\dtest$ records as $\phi_{err}$. We use DeepLift~\citep{shrikumar2017learning} from Captum library which recommends using a zero vector as a baseline.

\section{Evaluation}\label{sec:evaluation}
\setlength\tabcolsep{2.5pt}

\begin{wraptable}[22]{r}{0.5\textwidth}
\vspace{-2cm}
\caption{\textbf{Effectiveness of defenses.} For metrics, we use $\uparrow$ (resp. $\downarrow$) where higher (lower) value is better, and "\texttt{x}" is shorthand for $\phi^{\textbf{\texttt{(.)}}}_{(x)}$. (\phiacc is for context).}
\centering
\footnotesize
\resizebox{0.45\columnwidth}{!}{
\begin{tabular}{l | l | l | l}
\bottomrule

\toprule
\textbf{Defense} & \textbf{Metric} & \textbf{\texttt{FMNIST}} & \textbf{\texttt{UTKFACE}} \\
\midrule

\multirow{7}{*}{\textbf{No Defense}} & \textbf{u} ($\uparrow$) & 90.97 $\pm$ 0.18 & 80.28 $\pm$ 1.26\\ 
& \textbf{robacc} ($\uparrow$) & 7.96 $\pm$ 1.24 & 0.00 $\pm$ 0.00\\
& \textbf{ASR} ($\downarrow$) & 99.95 $\pm$ 0.04 & 99.98 $\pm$ 0.05\\
& \textbf{wmacc.Pre} ($\uparrow$) & 9.98 $\pm$ 0.28 & 0.00 $\pm$ 0.00\\
& \textbf{wmacc.In} ($\uparrow$) & 6.28 $\pm$ 1.20 & 62.21 $\pm$ 6.03\\
& \textbf{wmacc.Post} ($\uparrow$) & 0.00 $\pm$ 0.00 & 13.33 $\pm$ 6.32\\
& \textbf{RSD} ($\uparrow$) & 0.00 $\pm$ 0.00 & 0.00 $\pm$ 0.00  \\
& \textbf{eqodds} ($\downarrow$) & \cellcolor{gray!40} & 28.10 $\pm$ 6.34\\
& \textbf{dp} ($\downarrow$) & $\infty$ & $\infty$ \\
\midrule

\textbf{\defevasionIn} & \textbf{u} ($\uparrow$) & 86.93 $\pm$ 0.23 & 73.38 $\pm$ 1.15\\ 
& \textbf{robacc} ($\uparrow$) & 76.59 $\pm$ 0.28 & 37.38 $\pm$ 1.30\\
\midrule

\textbf{\defoutremIn} & \textbf{u} ($\uparrow$) & 89.38 $\pm$ 0.28 & 79.02 $\pm$ 0.30\\ 
& \textbf{ASR} ($\downarrow$) & 9.94 $\pm$ 0.24 &  56.62 $\pm$ 37.83\\
\midrule

\textbf{\defoutremPost} & \textbf{u} ($\uparrow$) & 86.48 $\pm$ 2.35 & 65.42 $\pm$ 3.27\\ 
& \textbf{ASR} ($\downarrow$) & 66.44 $\pm$ 21.30 & 8.59 $\pm$ 16.41\\
\midrule

\textbf{\defmodelwmPre} & \textbf{u} ($\uparrow$) & 90.15 $\pm$ 0.27 & 79.79 $\pm$ 0.39\\ 
& \textbf{wmacc} ($\uparrow$) & 99.91 $\pm$ 0.05 & 100.00 $\pm$ 0.00\\
\midrule

\textbf{\defmodelwmIn} & \textbf{u} ($\uparrow$) & 80.87 $\pm$ 0.88 & 66.71 $\pm$ 10.19\\ 
& \textbf{wmacc} ($\uparrow$) & 85.61 $\pm$ 2.50 & 93.74 $\pm$ 11.00\\
\midrule

\textbf{\defmodelwmPost} & \textbf{u} ($\uparrow$) & 90.56 $\pm$ 0.34 & 80.82 $\pm$ 0.45\\ 
& \textbf{wmacc} ($\uparrow$) & 100.00 $\pm$ 0.00 & 78.10 $\pm$ 9.33\\
\midrule

\textbf{\defdatawmPre} & \textbf{u} ($\uparrow$) & 90.31 $\pm$ 0.27 & 79.93 $\pm$ 0.37 \\ 
& \textbf{RSD} ($\uparrow$) & 100.00 $\pm$ 0.00 & 100.00 $\pm$ 0.00\\
\midrule

\textbf{\deffngprntPost}  & \textbf{u} ($\uparrow$) & No change & No change \\ 
& \textbf{pval} ($\downarrow$) & $<0.05$ & $<0.05$\\
\midrule

\textbf{\defdiffprivIn} & \textbf{u} ($\uparrow$) & 86.82 $\pm$ 0.11 & 74.07 $\pm$ 0.28\\ 
& \textbf{dp} ($\downarrow$) & $\epsilon_{dp}$ = 1.36 & $\epsilon_{dp}$ = 2.89\\
\midrule

\textbf{\deffairIn} & \textbf{u} ($\uparrow$) & \cellcolor{gray!40} & 76.85 $\pm$ 1.99\\ 
& \textbf{eqodds} ($\downarrow$) & \cellcolor{gray!40} & 10.89 $\pm$ 2.84\\
\midrule

\textbf{\defexplPost} & \textbf{u} ($\uparrow$) & No change & No change \\ 
& \textbf{err} ($\downarrow$) & 0.12 $\pm$ 0.03 & 0.59 $\pm$ 0.05\\
\bottomrule
 
\toprule
\end{tabular}
}
\label{tab:tab_eval_standalone}
\end{wraptable}

We evaluate individual defenses (\Snospace\ref{sec:standalone}), compare \method to na\"ive technique (\Snospace\ref{sec:evalPriorWork} and \Snospace\ref{sec:evalEmpirical}), study the impact of hyperparameter tuning (\Snospace\ref{sec:hyperparam}), and confirm that \method meets all requirements (\Snospace\ref{sec:otherReqs}).

\subsection{Evaluating Individual Defenses}\label{sec:standalone}

We evaluate the effectiveness of each defense by comparing the metrics $\phi^{\textbf{\texttt{D}}}_{(.)}$ to a ``no defense'' baseline.
We report the results in Table~\ref{tab:tab_eval_standalone}. We also report \phiacc to provide context but do not use it to evaluate accuracy of the technique.
We find that all the defense effectiveness metrics are better than the ``no defense'' baseline.
Once the defenses are applied, we use their respective $\phi^{\textbf{\texttt{D}}}_{(.)}$ as the ``single defense'' baseline to compare the effectiveness of the defense combinations later in \Snospace\ref{sec:evalEmpirical}.
For $\phi^{\text{\defexplPost}}_\text{err}$, we do not have a ``no defense'' baseline to compare with.
Assuming $\phi^{\text{\defexplPost}}_\text{err}$ is effective, we use it as the ``single defense'' baseline.





\begin{wraptable}[14]{r}{0.5\textwidth}
\vspace{-0.75cm}
    \caption{For each defense, we identify key parameters used in evaluating combination effectiveness (Figure~\ref{fig:flowchart}): type of change in \ref{step2} (Global, Local, or None); if defense uses a risk in \ref{step3} ("Yes" for backdoors or adversarial examples); and if defense protects against risk in \ref{step4}.}
    \centering
    \footnotesize
    \begin{tabular}{l|c|c|c}
    \bottomrule
    
    \toprule
       \textbf{Defense} & \ref{step2} & \ref{step3} & \ref{step4} \\
       \midrule
\textbf{\defevasionIn} & Global & No & Yes \\
\textbf{\defoutremIn} & Global & No & Yes \\
\textbf{\defoutremPost} & Global & No & Yes \\
\textbf{\defmodelwmPre} & Local & Yes & No \\
\textbf{\defmodelwmIn} & Global & Yes & No \\
\textbf{\defmodelwmPost} & Local & No & No \\
\textbf{\defdatawmPre} & Local & Yes & No \\
\textbf{\deffngprntPost} & None & No & No \\
\textbf{\defdiffprivIn} & Global & No & Yes \\
\textbf{\deffairIn} & Global & No & No \\
\textbf{\defexplPost} & None & No & No \\
    \bottomrule
    
    \toprule
    \end{tabular}
    \label{tab:decision_points}
\end{wraptable}

\subsection{Accuracy: using Prior Work}\label{sec:evalPriorWork}

Before empirically evaluating 48 defense combinations, we first identify the combinations which have been empirically evaluated in prior work (\Snospace\ref{sec:systematization} and Table~\ref{tab:tab_eval_combinations}). 
We identify eight combinations (\textbf{C1}-\textbf{C8}) whose results can be used as ground truth to compare the predictions of \method and the na\"ive technique (marked as \greenmark\xspace or \redmark\xspace in Table~\ref{tab:tab_eval_combinations}). We use \textcolor{ForestGreen}{green} and \textcolor{red}{red} to indicate alignment and conflict among defenses, respectively.
The prediction from a technique is accurate when \greenmarkOurs\xspace (or \redmarkOurs) for \method, or \greenmarkNaive\xspace (or \redmarkNaive) for na\"ive technique, match \greenmark\xspace (or \redmark) taken from prior work (\Snospace\ref{sec:systematization}) as ground truth.
We present additional details to make predictions in \ref{step2}, \ref{step3}, and \ref{step4} using \method in Table~\ref{tab:decision_points}.
\begin{itemize}[leftmargin=*]
\item \textbf{C1} \textbf{(\deffairPre + \defdiffprivPre)} can be combined effectively in the pre-training stage (\greenmark)~\citep{yaghini2023learning}. Na\"ive technique predicts \redmarkNaive\xspace (same stage) while \method predicts \greenmarkOurs\xspace (defenses make local changes in~\ref{step2}).

\item \textbf{C2} \textbf{(\defevasionIn +\deffngprntPost)} can be effectively combined (\greenmark)~\citep{szyller2023conflicting}. Na\"ive technique predicts \greenmarkNaive\xspace (different stages) while \method predicts \greenmarkOurs\xspace (\ref{step3}=no).

\item \textbf{C3} \textbf{(\defdiffprivIn + \deffngprntPost)} can be effectively combined (\greenmark)~\citep{szyller2023conflicting}. Na\"ive technique predicts \greenmarkNaive\xspace (different stages) while \method predicts \greenmarkOurs\xspace (\ref{step3}=no).

\item \textbf{C4} \textbf{(\defmodelwmPre +\defevasionIn)} are not effectively combined (\redmark)~\citep{szyller2023conflicting}. Na\"ive technique predicts \greenmarkNaive\xspace (different stages) while \method predicts \redmarkOurs\xspace (\defevasionIn makes poisons ineffective via unintended interaction in~\ref{step4}).

\item \textbf{C5} \textbf{(\defdatawmPre +\defevasionIn)} cannot be effectively combined (\redmark)~\citep{szyller2023conflicting}. Na\"ive technique predicts \greenmarkNaive\xspace (different stages) while \method predicts \redmarkOurs\xspace (\defevasionIn makes poisons ineffective via unintended interaction in~\ref{step4}).

\item \textbf{C6} \textbf{(\defmodelwmPre + \defdiffprivIn)} cannot be effectively combined (\redmark)~\citep{szyller2023conflicting}. Similar to \textbf{C5}, \method predictions this as \redmarkOurs\xspace while the na\"ive technique predicts \greenmarkNaive.

\item \textbf{C7} \textbf{(\defdatawmPre + \defdiffprivIn)} can be effectively combined (\greenmark)~\citep{szyller2023conflicting}. Na\"ive technique predicts \greenmarkNaive\xspace (different stages) while \method predicts \redmarkOurs\xspace (\defdiffprivIn reduces effectiveness of poisons via unintended interaction in~\ref{step4}).
Unlike backdoor-based watermarks used in our work, adversarial example-based watermarks used by Szyller and Asokan~\citep{szyller2023conflicting}, are inliers which are not suppressed by \defdiffprivIn.
Hence, \method's prediction differs.

\item \textbf{C8} \textbf{(\defdiffprivIn + \defexplPost)} can be effectively combined (\greenmark)~\citep{DPExplanations}.
Na\"ive technique predicts \greenmarkNaive\xspace (different stages) and \method predicts \greenmarkOurs\xspace (\ref{step3}=no).
\end{itemize}

\noindent\textbf{\emph{Of the eight combinations, \method predicts seven correctly, while na\"ive technique predicts four.
This gives a balanced accuracy of 90\% (TP=4, TN=3, FP=0, FN=1) for \method, and 40\% (TP=4, TN=0, FP=3, FN=1) for the na\"ive technique.}}


\subsection{Accuracy: via Empirical Evaluation}\label{sec:evalEmpirical}

We now empirically evaluate the remaining, previously unexplored, combinations to obtain the ground truth and then compute the accuracy of the predictions from both techniques.
After removing the eight combinations from prior work, we are left with 40 combinations. We also remove ten combinations where both defenses are applied during in-training. Here, both \method and the na\"ive technique predict \redmarkOurs\xspace and \redmarkNaive\xspace respectively.
To apply both defenses in the in-training phase, they can be modified for an effective combination (marked as \ref{t1:optimization} in \Snospace\ref{sec:framework}). However, this makes it invasive (violates \ref{n3:noninvasive}). Alternatively, defenses can be combined sequentially (e.g., pre-training on the first defense, then fine-tuning on the second), or by alternating the training of both defenses every few epochs. Since, these are non-standard approaches to apply existing defenses, we leave a comprehensive evaluation of ten combinations as future work.
Hence, \emph{\textbf{we are left with 30 combinations (C9-C38) for empirical evaluation.}}

\begin{table*}[!ht] 
\caption{\textbf{\underline{Evaluating combinations ($\hat{\texttt{D}}$):}} For brevity, "\texttt{x}"
 in \textbf{Metric} column is shorthand for $\phi^{\hat{\textbf{\texttt{D}}}}_{(x)}$. We use $\uparrow$ (resp. $\downarrow$) to indicate if a higher (lower) value is better. 
 For defense effectiveness, we use \colorbox{green!20}{green} when $\phi^{\hat{\textbf{\texttt{D}}}}_{(.)}$ is better or equal to ``single defense'' baseline; \colorbox{orange!20}{orange} for better or equal to ``single defense'' but worse than ``no defense''; \colorbox{red!20}{red} for worse than ``no defense''. For technique predictions, we use symbol \nocolorOurs\xspace (resp. \nocolorNaive) to refer to \method (na\"ive technique) and a color code \textcolor{ForestGreen}{green} (resp. \textcolor{red}{red}) to indicate alignment (conflict) among defenses. \defenseone and \defensetwo indicate order of applying defenses. (\phiacc is for context).}
\centering
\resizebox{0.95\textwidth}{!}{
\begin{tabular}{c| l | l | l | l || c | l | l | l | l}
\bottomrule

\toprule
& \textbf{Combinations} & \textbf{Metric} & \textbf{\texttt{FMNIST}} & \textbf{\texttt{UTKFACE}} & &\textbf{Combinations} & \textbf{Metric} & \textbf{\texttt{FMNIST}} & \textbf{\texttt{UTKFACE}}\\
\midrule

\multirow{3}{*}{\textbf{C9}} & \textbf{\defenseone: \defevasionIn} & \textbf{$\text{u}$} ($\uparrow$) & 90.38 $\pm$ 0.22 & 72.79 $\pm$ 0.53 & \multirow{3}{*}{\textbf{C24}} &
\textbf{\defenseone: \defmodelwmPre} & \textbf{$\text{u}$} ($\uparrow$) & 90.18 $\pm$ 0.21 & 79.76 $\pm$ 0.63\\ 
& \textbf{\defensetwo: \defmodelwmPost} & \textbf{$\text{wmacc}$} ($\uparrow$) & \cellcolor{green!20}100.00 $\pm$ 0.00 & \cellcolor{green!20}80.95 $\pm$ 7.13 &
& \textbf{\defensetwo: \defexplPost} & \textbf{$\text{err}$} ($\downarrow$) & \cellcolor{green!20}0.14 $\pm$ 0.04 & \cellcolor{green!20}0.02 $\pm$ 0.03\\
& \hfill (\greenmarkNaive, \greenmarkOurs)  & \textbf{${\text{robacc}}$} ($\downarrow$) & \cellcolor{green!20}80.43$\pm$0.85 & \cellcolor{green!20}42.00 $\pm$ 0.49 & & \hfill (\greenmarkNaive, \greenmarkOurs) & \textbf{${\text{wmacc}}$} ($\uparrow$) & \cellcolor{green!20}99.93 $\pm$ 0.06 & \cellcolor{green!20}99.96 $\pm$ 0.08\\
\midrule

\multirow{3}{*}{\textbf{C10}} & \textbf{\defenseone: \defoutremIn} & \textbf{$\text{u}$} ($\uparrow$) & 89.50 $\pm$ 0.21 & 79.25 $\pm$ 1.06 & \multirow{3}{*}{\textbf{C25}} & \textbf{\defenseone: \defmodelwmIn} & \textbf{$\text{u}$} ($\uparrow$) & 86.94 $\pm$ 0.50 & 72.16 $\pm$ 5.13 \\
& \textbf{\defensetwo: \deffngprntPost} & \textbf{$\text{ASR}$} ($\downarrow$) & \cellcolor{green!20}9.94 $\pm$ 0.22 & \cellcolor{green!20}56.09 $\pm$ 12.98 & & \textbf{\defensetwo: \defexplPost} & \textbf{$\text{err}$} ($\downarrow$) & \cellcolor{green!20}0.19 $\pm$ 0.07 & \cellcolor{green!20}0.37 $\pm$ 0.18\\
& \hfill (\greenmarkNaive, \greenmarkOurs) & \textbf{${\text{pval}}$} ($\downarrow$) & \cellcolor{green!20}<0.05 & \cellcolor{green!20}<0.05 & & \hfill (\greenmarkNaive, \greenmarkOurs)
& \textbf{${\text{wmacc}}$} ($\uparrow$) & \cellcolor{green!20}98.24 $\pm$ 0.66 & \cellcolor{green!20}97.60 $\pm$ 3.54\\
\midrule

\multirow{3}{*}{\textbf{C11}} & \textbf{\defenseone: \defoutremPost} & \textbf{$\text{u}$} ($\uparrow$) & 84.73 $\pm$ 1.72 & 63.70 $\pm$ 3.87 & \multirow{3}{*}{\textbf{C26}} & \textbf{\defenseone: \defdatawmPre} & \textbf{$\text{u}$} ($\uparrow$) & 90.04 $\pm$ 0.60 & 79.03 $\pm$ 1.10\\
& \textbf{\defensetwo: \deffngprntPost} & \textbf{$\text{ASR}$} ($\downarrow$) & \cellcolor{green!20}61.36 $\pm$ 23.96 & \cellcolor{green!20}0.02 $\pm$ 0.03 & & \textbf{\defensetwo: \defexplPost} & \textbf{$\text{err}$} ($\downarrow$) & \cellcolor{green!20}0.10 $\pm$ 0.04 & \cellcolor{green!20}0.54 $\pm$ 0.01\\
& \hfill (\redmarkNaive, \greenmarkOurs) & \textbf{${\text{pval}}$} ($\downarrow$) & \cellcolor{green!20}<0.05 & \cellcolor{green!20}<0.05 & & \hfill (\greenmarkNaive, \greenmarkOurs) & \textbf{${\text{RSD}}$} ($\uparrow$) & \cellcolor{green!20}100.00 $\pm$ 0.00 & \cellcolor{green!20}100.00 $\pm$ 0.00\\
\midrule

\multirow{3}{*}{\textbf{C12}} & \textbf{\defenseone: \defevasionIn} & \textbf{$\text{u}$} ($\uparrow$) & 87.10 $\pm$ 0.21 & 73.65 $\pm$ 1.21 & \multirow{3}{*}{\textbf{C27}} & \textbf{\defenseone: \defoutremIn} & \textbf{\textbf{$\text{u}$}} ($\uparrow$) & 89.39 $\pm$ 0.24 & 78.71 $\pm$ 0.20\\
& \textbf{\defensetwo: \defexplPost} & \textbf{$\text{err}$} ($\downarrow$) & \cellcolor{green!20}0.22 $\pm$ 0.01 & \cellcolor{green!20}0.15 $\pm$ 0.04 & & \textbf{\defensetwo: \defexplPost} & \textbf{$\text{ASR}$} ($\downarrow$) & \cellcolor{green!20}9.79 $\pm$ 0.15 & \cellcolor{green!20}44.35 $\pm$ 30.07\\
& \hfill (\greenmarkNaive, \greenmarkOurs) & \textbf{${\text{robacc}}$} ($\uparrow$) & \cellcolor{green!20}79.00 $\pm$ 0.21 & \cellcolor{green!20}39.27 $\pm$ 0.68 & & \hfill (\greenmarkNaive, \greenmarkOurs) & \textbf{\textbf{${\text{err}}$}} ($\downarrow$) & \cellcolor{green!20}0.06 $\pm$ 0.02 &  \cellcolor{green!20}0.47 $\pm$ 0.02\\
\midrule

\multirow{3}{*}{\textbf{C13}} & \textbf{\defenseone: \deffairIn} & \textbf{$\text{u}$} ($\uparrow$) & \cellcolor{gray!40} & 66.73 $\pm$ 3.24 & \multirow{3}{*}{\textbf{C28}} & \textbf{\defenseone: \defoutremPost} & \textbf{$\text{u}$} ($\uparrow$) & 84.62 $\pm$ 3.56 & 63.80 $\pm$ 3.37\\
& \textbf{\defensetwo: \defoutremPost} & \textbf{$\text{ASR}$} ($\downarrow$) & \cellcolor{gray!40} & \cellcolor{green!20}20.21 $\pm$ 39.90 & & \textbf{\defensetwo: \defexplPost} & \textbf{$\text{ASR}$} ($\downarrow$) & \cellcolor{green!20}76.11 $\pm$ 15.85 & \cellcolor{green!20}0.00 $\pm$ 0.00\\
& \hfill (\greenmarkNaive, \greenmarkOurs) & \textbf{${\text{eqodds}}$} ($\downarrow$) & \cellcolor{gray!40} & \cellcolor{green!20}2.72 $\pm$ 3.20 & & \hfill (\redmarkNaive, \greenmarkOurs) & \textbf{${\text{err}}$} ($\downarrow$) & \cellcolor{green!20}0.08 $\pm$ 0.01 & \cellcolor{green!20}0.15 $\pm$ 0.06\\
\midrule

\multirow{3}{*}{\textbf{C14}} & \textbf{\defenseone: \defmodelwmPre} & \textbf{$\text{u}$} ($\uparrow$) & \cellcolor{gray!40} & 79.02 $\pm$ 0.40 & \multirow{3}{*}{\textbf{C29}} & \textbf{\defenseone: \deffngprntPost} & \textbf{$\text{u}$} ($\uparrow$) & 90.56 $\pm$ 0.16 & 80.42 $\pm$ 0.59\\
& \textbf{\defensetwo: \deffairIn} & \textbf{$\text{wmacc}$} ($\uparrow$) & \cellcolor{gray!40} & \cellcolor{green!20}98.88 $\pm$ 2.13 & & \textbf{\defensetwo: \defexplPost} & \textbf{$\text{pval}$} ($\downarrow$) & \cellcolor{green!20}$<$0.05 & \cellcolor{green!20}$<$0.05\\
& \hfill (\greenmarkNaive, \greenmarkOurs) & \textbf{${\text{eqodds}}$} ($\downarrow$) & \cellcolor{gray!40} & \cellcolor{green!20}0.00 $\pm$ 0.00 & & \hfill (\redmarkNaive, \greenmarkOurs) & \textbf{${\text{err}}$} ($\downarrow$) & \cellcolor{green!20}0.11 $\pm$ 0.02 & \cellcolor{green!20}0.50 $\pm$ 0.03\\
\midrule

\multirow{3}{*}{\textbf{C15}} & \textbf{\defenseone: \deffairIn} & \textbf{$\text{u}$} ($\uparrow$) & \cellcolor{gray!40} & 76.95 $\pm$ 1.94 & \multirow{3}{*}{\textbf{C30}} & \textbf{\defenseone: \defdatawmPre} & \textbf{$\text{u}$} ($\uparrow$) & 90.19 $\pm$ 0.59 & 79.80 $\pm$ 0.48\\
& \textbf{\defensetwo: \defmodelwmPost} & \textbf{${\text{wmacc}}$} ($\uparrow$) & \cellcolor{gray!40} & \cellcolor{green!20}80.95 $\pm$ 0.00 & & \textbf{\defensetwo: \deffngprntPost} & \textbf{$\text{pval}$} ($\downarrow$) & \cellcolor{green!20}$<$0.05 & \cellcolor{green!20}$<$0.05\\
& \hfill (\greenmarkNaive, \greenmarkOurs) & \textbf{${\text{eqodds}}$} ($\downarrow$) & \cellcolor{gray!40} & \cellcolor{green!20}7.87 $\pm$ 4.72 & & \hfill (\greenmarkNaive, \greenmarkOurs) & \textbf{${\text{RSD}}$} ($\uparrow$) & \cellcolor{green!20}100.00 $\pm$ 0.00 & \cellcolor{green!20}100.00 $\pm$ 0.00\\
\midrule

\multirow{3}{*}{\textbf{C16}} & \textbf{\defenseone: \defdatawmPre} & \textbf{$\text{u}$} ($\uparrow$) & \cellcolor{gray!40} & 78.97 $\pm$ 1.21 & \multirow{3}{*}{\textbf{C31}} & \textbf{\defenseone: \defdiffprivIn} & \textbf{$\text{u}$} ($\uparrow$) & 86.83 $\pm$ 0.20 & 74.62 $\pm$ 0.49\\
& \textbf{\defensetwo: \deffairIn} & \textbf{RSD} ($\uparrow$) & \cellcolor{gray!40} & \cellcolor{green!20}100.00 $\pm$ 0.00 & & \textbf{\defensetwo: \defmodelwmPost} & \textbf{$\text{wmacc}$} ($\uparrow$) & \cellcolor{green!20}100.00 $\pm$ 0.00 & \cellcolor{green!20}79.05 $\pm$ 3.81\\
& \hfill (\greenmarkNaive, \greenmarkOurs) & \textbf{${\text{eqodds}}$} ($\downarrow$) & \cellcolor{gray!40} & \cellcolor{green!20}0.00 $\pm$ 0.00 & & \hfill (\greenmarkNaive, \greenmarkOurs) & ${\text{dp}}$ ($\downarrow$) & \cellcolor{green!20}$\epsilon$ = 1.36 &\cellcolor{green!20} $\epsilon$ = 2.89\\
\midrule

\multirow{3}{*}{\textbf{C17}} & \textbf{\defenseone: \deffairIn} & \textbf{$\text{u}$} ($\uparrow$) & \cellcolor{gray!40} & 78.67 $\pm$ 1.46 & \multirow{3}{*}{\textbf{C32}} & \textbf{\defenseone: \defdatawmPre} & \textbf{$\text{u}$} ($\uparrow$) & 90.24 $\pm$ 0.29 & 78.94 $\pm$ 0.95\\
& \textbf{\defensetwo: \deffngprntPost} & \textbf{$\text{pval}$} ($\downarrow$) & \cellcolor{gray!40} & \cellcolor{red!20}0.68 $\pm$ 0.21 & & \textbf{\defensetwo: \defmodelwmPost} & \textbf{$\text{RSD}$} ($\uparrow$) & \cellcolor{green!20}100.00 $\pm$ 0.00 & \cellcolor{green!20}100.00 $\pm$ 0.00\\
& \hfill (\greenmarkNaive, \greenmarkOurs) & \textbf{${\text{eqodds}}$} ($\downarrow$) & \cellcolor{gray!40} & \cellcolor{green!20}7.46 $\pm$ 5.43 & & \hfill (\greenmarkNaive, \greenmarkOurs) & \textbf{${\text{wmacc}}$} ($\uparrow$) & \cellcolor{green!20}100.00 $\pm$ 0.00 & \cellcolor{red!20}62.26 $\pm$ 3.77\\
\midrule

\multirow{3}{*}{\textbf{C18}} & \textbf{\defenseone: \deffairIn} & \textbf{$\text{u}$} ($\uparrow$) & \cellcolor{gray!40} & 80.52 $\pm$ 0.44 & \multirow{3}{*}{\textbf{C33}} & \textbf{\defenseone: \defoutremPost} & \textbf{$\text{u}$} ($\uparrow$) & 85.09 $\pm$ 1.94 & 67.09 $\pm$ 2.81\\
& \textbf{\defensetwo: \defexplPost} & \textbf{$\text{err}$} ($\downarrow$) & \cellcolor{gray!40} & \cellcolor{green!20}0.16 $\pm$ 0.06 & & \textbf{\defensetwo: \defmodelwmPost} & \textbf{$\text{wmacc}$} ($\uparrow$) & \cellcolor{green!20}100.00 $\pm$ 0.00 & \cellcolor{green!20}73.33 $\pm$ 8.83\\
& \hfill (\greenmarkNaive, \greenmarkOurs) & \textbf{${\text{eqodds}}$} ($\downarrow$) & \cellcolor{gray!40} & \cellcolor{green!20}12.62 $\pm$ 4.20 & & \hfill (\redmarkNaive, \greenmarkOurs) & \textbf{${\text{ASR}}$} ($\downarrow$) & \cellcolor{green!20}59.48 $\pm$ 24.91 & \cellcolor{green!20}40.20 $\pm$ 28.82\\
\midrule

\multirow{3}{*}{\textbf{C19}} & \textbf{\defenseone: \defoutremIn} & \textbf{$\text{u}$} ($\uparrow$) & 89.53 $\pm$ 0.36 & 79.00 $\pm$ 0.56 & \multirow{3}{*}{\textbf{C34}} & \textbf{\defenseone: \defdatawmPre} & \textbf{$\text{u}$} ($\uparrow$) & 90.31 $\pm$ 0.27 & 78.53 $\pm$ 1.75\\ 
& \textbf{\defensetwo: \defmodelwmPost} & \textbf{$\text{wmacc}$} ($\uparrow$) & \cellcolor{green!20}100.00 $\pm$ 0.00 & \cellcolor{green!20}69.52 $\pm$ 6.46 & & \textbf{\defensetwo: \defmodelwmPre} & \textbf{$\text{wmacc}$} ($\uparrow$) & \cellcolor{green!20}99.96 $\pm$ 0.0 & \cellcolor{green!20}100.00 $\pm$ 0.00\\
 & \hfill (\greenmarkNaive, \greenmarkOurs) & \textbf{${\text{ASR}}$} ($\downarrow$) & \cellcolor{green!20}10.48 $\pm$ 0.46 & \cellcolor{green!20}38.90 $\pm$ 38.73 & & \hfill (\redmarkNaive, \greenmarkOurs)  & \textbf{$\text{RSD}$} ($\uparrow$) & \cellcolor{green!20}100.00 $\pm$ 0.00 & \cellcolor{green!20}100.00 $\pm$ 0.00\\
\midrule

\multirow{3}{*}{\textbf{C20}} & \textbf{\defenseone: \defmodelwmPost} & \textbf{$\text{u}$} ($\uparrow$) & 90.93 $\pm$ 0.18 & 80.53 $\pm$ 0.23 & \multirow{3}{*}{\textbf{C35}} & \textbf{\defenseone: \defevasionIn} &  \textbf{$\text{u}$} ($\uparrow$) & 90.39 $\pm$ 0.63 & 80.28 $\pm$ 0.39\\ 
& \textbf{\defensetwo: \defexplPost} & \textbf{${\text{wmacc}}$} ($\uparrow$) & \cellcolor{green!20}100.00 $\pm$ 0.00 & \cellcolor{green!20}72.38 $\pm$ 3.56 & & \textbf{\defensetwo: \defoutremPost} & \textbf{${\text{robacc}}$} ($\uparrow$) & \cellcolor{orange!20}46.00 $\pm$ 1.02 & \cellcolor{red!20}0.00 $\pm$ 0.00\\
& \hfill (\redmarkNaive, \greenmarkOurs) & \textbf{$\text{err}$} ($\downarrow$) &  \cellcolor{green!20}0.11 $\pm$ 0.02 & \cellcolor{green!20}0.55 $\pm$ 0.02 & & \hfill (\greenmarkNaive, \greenmarkOurs) & \textbf{$\text{ASR}$} ($\downarrow$) & \cellcolor{orange!20}79.68 $\pm$ 10.25 & \cellcolor{green!20}0.00 $\pm$ 0.00\\
\midrule

\multirow{3}{*}{\textbf{C21}} & \textbf{\defenseone: \defdatawmPre} & \textbf{$\text{u}$} ($\uparrow$) & 89.46 $\pm$ 0.32 & 79.00 $\pm$ 0.67 & \multirow{3}{*}{\textbf{C36}} & \textbf{\defenseone: \defmodelwmPre} & \textbf{$\text{u}$} ($\uparrow$) & 89.48 $\pm$ 0.15  & 79.20 $\pm$ 0.60\\ 
& \defensetwo: \defoutremIn & \textbf{$\text{ASR}$} ($\downarrow$) & \cellcolor{green!20}10.18 $\pm$ 0.40 & \cellcolor{orange!20}77.39 $\pm$ 35.23 & & \textbf{\defensetwo: \defoutremIn}  & \textbf{$\text{ASR}$} ($\downarrow$) & \cellcolor{green!20}10.18 $\pm$ 0.46 & \cellcolor{green!20}46.92 $\pm$ 36.92\\
& \hfill (\greenmarkNaive, \redmarkOurs) & \textbf{$\text{RSD}$} ($\uparrow$) & \cellcolor{red!20}0.00 $\pm$ 0.00 & \cellcolor{orange!20}80.00 $\pm$ 40.00 & & \hfill (\greenmarkNaive, \redmarkOurs) & \textbf{${\text{wmacc}}$} ($\uparrow$) & \cellcolor{red!20}10.18 $\pm$ 0.46 & \cellcolor{orange!20}46.92 $\pm$ 36.92 \\
\midrule

\multirow{3}{*}{\textbf{C22}} & \textbf{\defenseone: \defdatawmPre} & \textbf{$\text{u}$} ($\uparrow$) & 84.45 $\pm$ 0.56 & 79.88 $\pm$ 0.27 & \multirow{3}{*}{\textbf{C37}} & \textbf{\defenseone: \defmodelwmPre} & \textbf{$\text{u}$} ($\uparrow$) &  82.86 $\pm$ 4.16 & 64.09 $\pm$ 3.09\\ 
& \textbf{\defensetwo: \defmodelwmIn} & \textbf{${\text{wmacc}}$} ($\uparrow$) & \cellcolor{green!20}89.25 $\pm$ 3.48 & \cellcolor{green!20}99.98 $\pm$ 0.03 & & \textbf{\defensetwo: \defoutremPost}  & \textbf{$\text{ASR}$} ($\downarrow$) & \cellcolor{green!20}71.32 $\pm$ 14.11 & \cellcolor{green!20}0.00 $\pm$ 0.00\\
& \hfill (\greenmarkNaive, \greenmarkOurs) & \textbf{$\text{RSD}$} ($\uparrow$) & \cellcolor{green!20}100.00 $\pm$ 0.00 & \cellcolor{green!20}100.00 $\pm$ 0.00 & & \hfill (\greenmarkNaive, \redmarkOurs) & \textbf{${\text{wmacc}}$} ($\uparrow$) & \cellcolor{orange!20}71.31 $\pm$ 14.10 & \cellcolor{red!20}0.00 $\pm$ 0.00\\
\midrule

\multirow{3}{*}{\textbf{C23}} & \textbf{\defenseone: \defdatawmPre} & \textbf{$\text{u}$} ($\uparrow$) & 82.90 $\pm$ 2.06 & 69.02 $\pm$ 1.96 & \multirow{3}{*}{\textbf{C38}} & \textbf{\defenseone: \defmodelwmIn} & \textbf{$\text{u}$} ($\uparrow$) & 66.68 $\pm$ 9.80  & 73.69 $\pm$ 3.01\\ 
& \textbf{\defensetwo: \defoutremPost} & \textbf{$\text{ASR}$} ($\downarrow$) & \cellcolor{green!20}64.55 $\pm$ 21.23 & \cellcolor{green!20}0.01 $\pm$ 0.01 & & \textbf{\defensetwo: \defoutremPost} & \textbf{$\text{ASR}$} ($\downarrow$) & \cellcolor{orange!20}58.59 $\pm$ 19.22 & \cellcolor{red!20}99.60 $\pm$ 0.37\\
& \hfill (\greenmarkNaive, \redmarkOurs) & \textbf{$\text{RSD}$} ($\uparrow$) & \cellcolor{orange!20}80.00 $\pm$ 40.00 & \cellcolor{orange!20}20.00 $\pm$ 40.00 & & \hfill (\greenmarkNaive, \redmarkOurs) & \textbf{${\text{wmacc}}$} ($\uparrow$)  & \cellcolor{orange!20}58.65 $\pm$ 19.23 & \cellcolor{green!20}99.73 $\pm$ 0.29\\
\bottomrule

\toprule
\end{tabular}
}
\label{tab:tab_eval_combinations}
\end{table*}

\noindent\textbf{\ul{Predictions from Techniques.}} Before evaluating 30 combinations, we denote the defenses as \defenseone and \defensetwo based on the order in which they are applied.
We obtain predictions from \method and the na\"ive techniques, and indicate them as a tuple: (Na\"ive prediction, \method prediction).
These are indicated in Table~\ref{tab:tab_eval_combinations}.
We use the information in Table~\ref{tab:decision_points} to make predictions in \ref{step2}-\ref{step4} for \method.
\begin{itemize}[leftmargin=*]
\item For defenses applied in the same stage (\ref{step1}=yes), the na\"ive technique predicts \redmarkNaive.
We have the following cases to determine the prediction from \method:
\begin{enumerate}[leftmargin=*]
    \item \defensetwo makes local/no changes (\ref{step2}=no), \method predicts this as \greenmarkNaive. We mark them as (\redmarkNaive, \greenmarkOurs) which include \textbf{C11}, \textbf{C20}, \textbf{C28}, \textbf{C29}, \textbf{C33}, and \textbf{C34}.
    \item \defensetwo makes global changes (\ref{step2}=yes), and \method predicts this as \redmarkOurs. We mark them as (\redmarkNaive, \redmarkOurs) but we did not observe any such combinations.
\end{enumerate}

\item For defenses applied in different stages (\ref{step1}=no), the na\"ive technique predicts \greenmarkNaive.
We have the following cases to determine the prediction from \method:
\begin{enumerate}[leftmargin=*]
\item \defenseone does not use a risk (\ref{step3}=no) and hence, \defenseone and \defensetwo do not conflict. We mark them as (\greenmarkNaive, \greenmarkOurs): \textbf{C9}, \textbf{C10}, \textbf{C12}, \textbf{C13}, \textbf{C15}, \textbf{C17}-\textbf{C19}, \textbf{C27}, \textbf{C31}, and \textbf{C35}.

\item \defenseone, such as \defmodelwmPre and \defdatawmPre, uses a risk (\ref{step3}=yes), but \defensetwo does not protect against this risk (\ref{step4}=no). Hence, there is no conflict and we mark such combinations as (\greenmarkNaive, \greenmarkOurs) which include \textbf{C14}, \textbf{C16}, \textbf{C22}, \textbf{C24}-\textbf{C26}, \textbf{C30}, and \textbf{C32}.

\item \defenseone, such as \defmodelwmPre and \defdatawmPre, uses a risk (\ref{step3}=yes), and \defensetwo mitigates these risks (e.g., \defoutrem). There is a conflict and we mark them as (\greenmarkNaive, \redmarkOurs) which include \textbf{C21}, \textbf{C23}, \textbf{C36}-\textbf{C38}.
\end{enumerate}
\end{itemize}
We evaluate the 30 combinations on \texttt{FMNIST} and \texttt{UTKFACE} (Table~\ref{tab:tab_eval_combinations}). 
For each combination, we compare the effectiveness metrics for each defense to ``single defense'' from Table~\ref{tab:tab_eval_standalone}.
We use \colorbox{green!20}{green} to indicate that the metrics are better or similar to ``single defense''; \colorbox{orange!20}{orange} for worse than single defense but better than ``no defense''; and \colorbox{red!20}{red} for similar or worse than ``no defense''.
Metrics marked as \colorbox{orange!20}{orange} can still be useful since it provides some protection compared to ``no defense''.
We consider the worst case by a treating a combination as a conflict if atleast one dataset has atleast one metric marked as \colorbox{orange!20}{orange} or \colorbox{red!20}{red}.

\noindent\textbf{\emph{Of the 30 combinations, \method predicts 27 correctly, while the na\"ive method predicts only 18.
This gives a balanced accuracy of 81\% (TP=22, TN=5, FP=3, and FN=0) for \method compared to 36\% (TP=16, TN=0, FP=8, and FN=6) for the na\"ive technique.}}

\begin{takeaway}
\textbf{Takeaway:} By explicitly accounting for reasons underlying conflicts among defenses, \method achieves higher accuracy than the na\"ive technique (satisfies \ref{n1:effective}).
\end{takeaway}

\subsection{Hyperparameter Tuning for Combinations}\label{sec:hyperparam}

\begin{wraptable}[14]{r}{0.5\textwidth}
\vspace{-1.5cm}
\centering
\footnotesize
\caption{Configurations for hyperparameter tuning of defenses in conflicting combinations.}
\resizebox{0.5\columnwidth}{!}{
\begin{tabular}{l|l|l}
\bottomrule

\toprule
\textbf{Defense} & \textbf{Hyperparameter} & \textbf{Values} \\
\bottomrule

\toprule
\multirow{1}{*}{\textbf{\defevasionIn}} & Regularization & \{2, 4, 6 (default), 8\} \\
\midrule
\multirow{1}{*}{\textbf{\defoutremPost}} & Pruning threshold &  0.6–1.5 (step 0.05) \\
\midrule
\multirow{2}{*}{\textbf{\defmodelwmPre}} & Trigger size & \{3$\times$3, 5$\times$5 (default)\} \\
 & Watermark fraction & \{0.1 (default), 0.2, 0.3\} \\
\midrule
\multirow{3}{*}{\textbf{\defmodelwmIn}} & Watermark fraction & \{0.1, 0.2, 0.3\} \\
 & Training noise & \{0.5, 0.75, 1.0 (default), 1.25\} \\
 & Noise step size & \{0.05 (default), 0.10, 0.15\} \\
\midrule
\multirow{1}{*}{\textbf{\defmodelwmPost}} & Watermark fraction & \{0.002 (default), 0.01, 0.02\} \\
\midrule
\multirow{2}{*}{\textbf{\defdatawmPre}} & Trigger size & \{3$\times$3, 5$\times$5 (default)\} \\
 & Watermark fraction & \{0.1 (default), 0.2, 0.3\} \\
\midrule
\multirow{1}{*}{\textbf{\deffairIn}} & Regularization & \{0.5, 1 (default), 1.5, 2\} \\
\midrule
\multirow{2}{*}{\textbf{\deffngprntPost}} & Iterations & \{25, 50 (default), 75, 100\} \\
 & \# Fingerprints & \{100 (default), 150, 200\} \\
\bottomrule

\toprule
\end{tabular}
}
\label{tab:hyperparams}
\end{wraptable}
We check if hyperparameter tuning can resolve conflicts to see if it can turn
\begin{enumerate*}[label={(\roman*)}]
\item \textbf{false positives} (predicted as aligned, but empirically conflicting) into \textbf{true positives}, and
\item \textbf{true negatives} (predicted and confirmed as conflict) into \textbf{false negatives}.
\end{enumerate*}
We exclude \textbf{false negatives} (not observed in our evaluation), and \textbf{true positives}, correctly predicted and confirmed as aligned (cannot be improved further with hyperparameter tuning).
We use grid search and identify various hyperparameter configurations for defenses in conflicting combinations (Table~\ref{tab:hyperparams}).

\noindent\textbf{\underline{Do False Positives turn to True Positives?}} This includes three combinations (\textbf{C17}, \textbf{C32}, \textbf{C35}), and helps investigate \method errors.
\begin{itemize}[leftmargin=*]
\item \noindent\textbf{C17} \textbf{(\deffairIn + \deffngprntPost)}. We empirically observe a conflict as \deffngprntPost is ineffective ($\phi_{\text{pval}}$ > $0.05$), and \method incorrectly predicts the combination as \greenmarkOurs\xspace in \ref{step3}.
We explore the following hyperparameters: regularization for \deffairIn, iterations, and number of fingerprints for \deffngprntPost.
For each dataset, we have 36 experiments (= 4 $\times$ 4 $\times$ 3). None of the experiments alleviated the conflict.
Following prior work~\citep{szyller2023conflicting}, we speculate that \deffngprntPost is ineffective because it relies on the decision boundary, which shifts significantly after applying \deffairIn.

\item \noindent\textbf{C32} \textbf{(\defdatawmPre + \defmodelwmPost)}. The combination is empirically effective for \texttt{FMNIST} but not for \texttt{UTKFACE} where $\phi_{wmacc}$ is less than the ``single defense'' baseline. 
\method incorrectly predicts this combination as \greenmarkOurs\xspace in \ref{step4}.
We explore the following hyperparameters:
\begin{enumerate*}[label={(\roman*)},itemjoin={;\xspace}]
\item trigger size and watermark fraction for \defdatawmPre
\item watermark fraction for \defmodelwmPost.
\end{enumerate*}
For \texttt{UTKFACE}, we evaluate 18 experiments (= 2 $\times$ 3 $\times$ 3).
One experiment with 3$\times$3 trigger size (\defdatawmPre), 30\% watermarks (\defdatawmPre), and 2\% watermarks (\defmodelwmPost), we get
\phiacc = 75.98 $\pm$ 0.61, $\phi_{RSD}$ = 100.00 $\pm$ 0.00 (\colorbox{green!20}{green}), and $\phi_{wmacc}$ = 70.19 $\pm$ 4.61 (\colorbox{green!20}{green}).
Hence, we remove the conflict and the false positive.

\item\noindent\textbf{C35} \textbf{(\defevasionIn + \defoutremPost)}. Empirically, there is a conflict as \defevasionIn is ineffective (poor $\phi_{robacc}$), and \method incorrectly predicts as \greenmarkOurs\xspace in \ref{step4}.
We vary the regularization hyperparameter (\defevasionIn), and pruning thresholds (\defoutremPost).
None of the experiments removed the conflict.
We speculate that the model parameters memorizing poisons and adversarial examples overlap.
Thus, pruning a model (to reduce $\phi_{ASR}$) trained with \defevasionIn, also reduces $\phi_{robacc}$, resulting in a conflict.
\end{itemize}

\noindent\textbf{\underline{Do True Negatives turn to False Negatives?}} We evaluate hyperparameter tuning for five combinations (\textbf{C21}, \textbf{C23}, \textbf{C36}, \textbf{C37}, and \textbf{C38}).
\begin{itemize}[leftmargin=*]
\item \textbf{C21} \textbf{(\defdatawmPre + \defoutremIn)}. We consider trigger size, and watermark fraction for \defdatawmPre.
For each dataset, we get six experiments (=2$\times$3).
None of them removed the conflict since \defoutremIn mitigates backdoors for \defdatawmPre.

\item\noindent\textbf{C23} \textbf{(\defdatawmPre + \defoutremPost)}.
We tune the same hyperparameters for \defdatawmPre as in \textbf{C21}.
For \defoutremPost, we sweep across various pruning thresholds.
For each dataset, we get six experiments (= 2 $\times$ 3).
For \texttt{FMNIST}, trigger size of 3$\times$3 and 30\% watermarks, gives \phiacc=82.00 $\pm$ 5.50; $\phi_{RSD}$= 100.00 $\pm$ 0.00; $\phi_{ASR}$= 59.90 $\pm$ 12.81. This is marked as no conflict (\colorbox{green!20}{green}). However, there is a conflict for \texttt{UTKFACE}, making the overall combination a conflict.

\item \noindent\textbf{C36} \textbf{(\defmodelwmPre + \defoutremIn)}.
We tune the same hyperparameters for \defmodelwmPre, as \defdatawmPre in \textbf{C21}.
For each dataset, we have six experiments (=2$\times$3).
None of them removed the conflict since \defoutremIn mitigates backdoors for \defmodelwmPre.

\item\noindent\textbf{C37} \textbf{(\defmodelwmPre + \defoutremPost)}.
We tune the same hyperparameters for \defmodelwmPre, as \defdatawmPre in \textbf{C21}.
For \defoutremPost, we sweep across various pruning thresholds.
For each dataset, we get six experiments (=2$\times$3).
None of them removed the conflict since \defoutremPost mitigates backdoors for \defmodelwmPre.

\item\noindent\textbf{C38} \textbf{(\defmodelwmIn + \defoutremPost)}.
We tune the fraction of watermarks, training noise, and step size for \defmodelwmIn.
For \defoutremPost, we sweep across various pruning thresholds.
For each dataset, we have 36 experiments (=3$\times$4$\times$3).
None of them removed the conflict since \defoutremPost mitigates the backdoors for \defmodelwmIn.
\end{itemize}

\noindent\textbf{\underline{Summary.}} 
We find that hyperparameter tuning is useful in two combinations (\textbf{C23} and \textbf{C32}). 
For \textbf{C32}, we removed the false positive, thereby increasing 
\method's balanced accuracy to 86\% (from 81\%).
For \textbf{C23}, we could remove conflict for one of the two datasets, but the combination was still marked as a conflict (no additional false negatives).   

\begin{takeaway}
\textbf{Takeaway:} Hyperparameter tuning for \emph{conflicting combinations} is important to check if it turns (a) false positives to true positives, or (b) true negatives to false negatives.
\end{takeaway}

\change{\noindent\textbf{\underline{Factors for Hyperparameter Tuning Effectiveness.}}
We identify factors affecting the effectiveness of hyperparameter tuning and highlight how these factors apply to some conflicting combinations:
\begin{itemize}[leftmargin=*]
    \item \textbf{Expressiveness of hyperparameters:} Tuning is only effective if there are hyperparameters that directly influence the conflicting interaction. If the interaction is insensitive to changes in a hyperparameter, tuning will not resolve the conflict. This could be the reason for hyperparameter tuning being ineffective for some combinations (\textbf{C21}, \textbf{C36}, \textbf{C37}, and \textbf{C38}).
    \item \textbf{Search Space:} If the search space is narrow, tuning may never find the optimal configuration to resolve conflicts. Conversely, a sufficiently broad search space increases the chance of finding a configuration that decouples the objectives. There is a possibility that tuning did not work for some combinations as our search space was not broad enough (e.g., \textbf{C17}, \textbf{C32}, \textbf{C35}).
    \item \textbf{Fundamental Incompatibility:} Some defenses are fundamentally incompatible and cannot be resolved by tuning. We discuss this further below.
    \item \textbf{Optimization Landscape:} If the loss landscape contains many local minima, tuning may not find the optimal configuration to resolve conflicts. A practitioner can try more sophisticated tuning instead of grid search (e.g., randomized search or Bayesian optimization), to see if it resolves conflicts.
\end{itemize}

\noindent\textbf{\underline{Limitations.}}
When tuning fails to resolve a conflict, we cannot conclusively mark the combination as a conflict: despite considering a feasible search space in our evaluation, some conflict-resolving hyperparameters may have been missed given the vast number of possibilities.
Also, when a combination is marked as conflict (e.g., ``\ref{step4}=Yes'' for \textbf{C21}, \textbf{C36}, \textbf{C37}, and \textbf{C38}), it suggests an \emph{empirical incompatibility} among defenses but does not imply a conclusive proof or \emph{fundamental incompatibility}. Establishing fundamental incompatibility requires a theoretical analysis which is left as future work.
}

\subsection{Other Requirements}\label{sec:otherReqs}

Having shown that the na\"ive technique does not perform as well as \method, we discuss how \method meets the remaining requirements: scalability (\ref{n2:scalable}), non-invasive (\ref{n3:noninvasive}), and generality (\ref{n4:general}).

\setlength\tabcolsep{2pt}
\begin{wraptable}[22]{r}{0.5\textwidth}
\vspace{-0.85cm}
\caption{\textbf{Scalability (\ref{n2:scalable}) of \method to \defenseone, \defensetwo, and \defensethree (in order).}
Color coding and notations are same as in Table~\ref{tab:tab_eval_combinations}.
}
\centering
\footnotesize
\resizebox{0.45\columnwidth}{!}{
\begin{tabular}{c | l | l | l | l }
\bottomrule

\toprule
& \textbf{Combinations} & \textbf{Metric} & \textbf{\texttt{FMNIST}} & \textbf{\texttt{UTKFACE}} \\
\midrule


\multirow{4}{*}{\textbf{C39}} & \defenseone: \defevasionIn & $\text{u}$ ($\uparrow$) & 87.38 $\pm$ 0.15 & 74.34 $\pm$ 0.72\\
& \defensetwo: \defexplPost & $\text{robacc}$ ($\uparrow$) & \cellcolor{green!20}79.37 $\pm$ 0.29 & \cellcolor{green!20}39.21 $\pm$ 0.32\\
& \defensethree: \defmodelwmPost & $\text{err}$ ($\downarrow$) & \cellcolor{green!20}0.96 $\pm$ 0.14 & \cellcolor{green!20}0.17 $\pm$ 0.05\\
& & $\text{wmacc}$ ($\uparrow$) & \cellcolor{green!20}100.00 $\pm$ 0.00 & \cellcolor{green!20}73.33 $\pm$ 8.83\\
\midrule

\multirow{4}{*}{\textbf{C40}} & \defenseone: \defoutremIn & $\text{u}$ ($\uparrow$) & 89.47 $\pm$ 0.24 & 79.42 $\pm$ 0.51 \\
& \defensetwo: \defexplPost & $\text{ASR}$ ($\downarrow$) & \cellcolor{green!20}9.81 $\pm$ 0.12 & \cellcolor{green!20}66.74 $\pm$ 12.11  \\
& \defensethree: \defmodelwmPost & $\text{err}$ ($\downarrow$) & \cellcolor{green!20}0.06 $\pm$ 0.02 & \cellcolor{green!20}0.52 $\pm$ 0.04  \\
& & $\text{wmacc}$ ($\uparrow$) & \cellcolor{green!20}100.00 $\pm$ 0.00 & \cellcolor{green!20}77.14 $\pm$ 11.82 \\
\midrule

\multirow{4}{*}{\textbf{C41}} & \defenseone: \defoutremPost & $\text{u}$ ($\uparrow$) & 89.47 $\pm$ 0.24 & 67.04 $\pm$ 3.35 \\
& \defensetwo: \defexplPost & $\text{ASR}$ ($\downarrow$) & \cellcolor{green!20}9.81 $\pm$ 0.12 & \cellcolor{green!20}1.85 $\pm$ 3.39\\
& \defensethree: \defmodelwmPost & $\text{err}$ ($\downarrow$) & \cellcolor{green!20}0.06 $\pm$ 0.02 & \cellcolor{green!20}0.17 $\pm$ 0.10\\
& & $\text{wmacc}$ ($\uparrow$) & \cellcolor{green!20}100.00 $\pm$ 0.00 & \cellcolor{green!20}81.90 $\pm$ 7.00\\
\midrule

\multirow{4}{*}{\textbf{C42}} & \defenseone: \defdatawmPre & $\text{u}$ ($\uparrow$) & \cellcolor{gray!40} & 77.53 $\pm$ 1.75\\ 
& \defensetwo: \deffairIn & $\text{wmacc}$ ($\uparrow$) & \cellcolor{gray!40} & \cellcolor{green!20}100.00 $\pm$ 0.00\\
& \defensethree: \defexplPost & $\text{eqodds}$ ($\downarrow$) & \cellcolor{gray!40} & \cellcolor{green!20}0.00 $\pm$ 0.00\\
& & $\text{err}$ ($\downarrow$) & \cellcolor{gray!40} &\cellcolor{green!20}0.01 $\pm$ 0.00\\
\midrule

\multirow{4}{*}{\textbf{C43}} & \defenseone: \defdatawmPre & $\text{u}$ ($\uparrow$) & \cellcolor{gray!40} & 79.17 $\pm$ 0.93 \\
& \defensetwo: \deffairIn & $\text{RSD}$ ($\uparrow$) & \cellcolor{gray!40} & \cellcolor{green!20}100.00 $\pm$ 0.00 \\
& \defensethree: \defmodelwmPost & $\text{eqodds}$ ($\downarrow$) & \cellcolor{gray!40} & \cellcolor{green!20}0.00 $\pm$ 0.00 \\
& & $\text{wmacc}$ ($\uparrow$) & \cellcolor{gray!40} & \cellcolor{green!20}73.33 $\pm$ 7.12 \\
\midrule


\multirow{4}{*}{\textbf{C44}} & \defenseone: \deffairIn & $\text{u}$ ($\uparrow$) & \cellcolor{gray!40} & 69.42 $\pm$ 2.09\\
& \defensetwo: \defoutremPost & $\text{eqodds}$ ($\downarrow$) & \cellcolor{gray!40} & \cellcolor{green!20}8.12 $\pm$ 4.49\\
& \defensethree: \defexplPost & $\text{ASR}$ ($\downarrow$) & \cellcolor{gray!40} & \cellcolor{green!20}0.13 $\pm$ 0.25\\
& & $\text{err}$ ($\downarrow$) & \cellcolor{gray!40} & \cellcolor{green!20}0.05 $\pm$ 0.02\\


\bottomrule

\toprule
\end{tabular}
}
\label{tab:tab_eval_scalable}
\end{wraptable}
\noindent\textbf{\underline{Scalability (\ref{n2:scalable}).}} None of the prior works have considered more than two defenses. Since \method allows for applying defenses in three stages of the ML pipeline, it should theoretically support at least three defenses. To illustrate this, we follow the instructions in \Snospace\ref{sec:approach} to extend \method beyond two defenses. 
We begin with pairwise combinations predicted as effective (marked as \greenmarkOurs\xspace in Table~\ref{tab:tab_eval_combinations}), which align with empirical evaluation, and then include additional defenses.
We consider five combinations with three defenses each, which should be effectively combines (marked as \greenmarkOurs). 
We report the results in Table~\ref{tab:tab_eval_scalable} and find that it is indeed possible to effectively combine three defenses using \method.
\emph{Overall, \method scales to more than two defenses (\ref{n2:scalable}).}
\change{These are illustrative examples to show that \method is scalable to more than two defenses. Our goal was not an exhaustive evaluation of the large number of multi-way combinations, but only to show that effective multi-way combinations--previously unexplored--are possible. A comprehensive evaluation to identify false positives and negatives, is left as future work.}

\noindent\textbf{\underline{Non-Invasive (\ref{n3:noninvasive}).}}
\method extends \ref{t2:intervention} and hence, inherits the non-invasive requirement. 
We use existing defenses proposed in the literature without modifying them, and only adapting them to our datasets.
\noindent\emph{In summary, \method satisfies \ref{n3:noninvasive}.}

\noindent\textbf{\underline{General (\ref{n4:general}).}}
\change{\method identifies a conflict based on (a) relative position of the defenses in ML pipeline, and (b) the mechanisms that underlie them (whether one defense uses a risk that is being protected by a later defense). \method does not rely on specific defenses. Hence, \method is likely to be applicable beyond the initial set of defenses we used for evaluating our work (commonly available defenses from the literature).}
%
\change{For identifying conflicts of a new defense with others, a practitioner can first identify its position in the ML pipeline, and determine whether it uses a risk defended by a later defense. Since these are independent of any specific defense, \method is applicable for any defense that we can map to \method's flowchart (Figure~\ref{fig:flowchart}).
Furthermore, we select specific defense implementations based on their availability (see \Snospace\ref{sec:metrics}). However, other implementations can be used and should not effect our conclusions. 
}

\begin{takeaway}
\textbf{Takeaway:} \method scales beyond two defenses (\ref{n2:scalable}), is non-invasive (\ref{n3:noninvasive}), and general (\ref{n4:general}).
\end{takeaway}
\section{Discussion, Conclusions, and Future Work}\label{sec:discussions}

\noindent\textbf{\underline{Note on Model Utility.}}
So far, we have focused only on \emph{effectiveness}, examining how combining defenses impacts the effectiveness of each individual defense. 
An additional pre-requisite for deploying a defense combination is whether it negatively impacts \emph{model utility}.
We can define a defense combination to be \emph{viable} if it is (a) effective and (b) incurs only a minimal utility drop compared to lowest of the ``single defense'' baseline.
In Table~\ref{tab:tab_eval_combinations}, we observe that all the combinations which \method predicted as effective are also viable. For \textbf{C15}, \textbf{C27}, and \textbf{C30}, the utility is worse than the ``single defense'' baseline. These were already flagged as ineffective.
We did not observe any combinations which are effective but not viable.

Extending \method to predict the viability of the combinations is challenging, since it is unclear how a defense impacts model utility.
\change{For instance, there can be the following two cases:
\begin{itemize}[leftmargin=*]
    \item For both defenses, if utility is either better or similar to the ``single defense'' baseline, it is likely that the combination will have acceptable utility.
    \item If the utility degrades for both defenses, the combination is likely to have poor utility and hence non-viable. However, it is also possible (as seen in Table~\ref{tab:tab_eval_combinations}) that the utility of the combination does not fall below the ``minimum utility for single defenses'' baseline, even if some, but not all, constituent defenses fall below their respective ``no-defense'' baseline. It is unclear what mechanisms account for this phenomenon. 
\end{itemize}
Hence, quantifying the impact of individual defenses on utility is an open problem for several defenses (e.g., adversarial training~\citep{trades,tsipras2018robustness,NEURIPS2020_61d77652,pang2022robustness,raghunathan2020understanding} and differential privacy~\citep{dpacc,NEURIPS2023_1165af8b,papernot2021tempered,Tramr2020DifferentiallyPL}), and an area of active research.}
\textit{Viability} can be included as a requirement in \Snospace\ref{sec:desiderata}, and extending \method for viable combinations is left as future work. 


\change{\noindent\textbf{\underline{Practical Considerations.}} We discuss the impact on other considerations such as computational cost and latency. The computational cost (for training) and the latency (for inference) are the sum of the costs and latencies, incurred by the constituent defenses when applied individually.
\begin{itemize}[leftmargin=*]
    \item Pre-training, in-training, and some post-training (e.g., pruning a model) defenses incur a reasonable one-time cost, assuming the practitioner has basic resources to train ML models (e.g., GPUs). These do not have any impact on the inference-time latency.
    \item For some post-training defenses, there is no training cost but incur a per-inference latency for transforming the inputs or outputs.
\end{itemize}
}

\change{\noindent\textbf{\underline{Real-world Impact.}} Following the experimental setup in prior work, we evaluated \method in a lab setting and not on real-world models, which was not realistic for our experiments.
The real-world impact of successful combinations depends on the individual application. Practitioners have to decide which combination should be applied for a given application: for instance, in credit card approval, robustness, fairness, and explainability are important properties, while in medical diagnosis, privacy may also be essential.}

\change{\noindent\textbf{\underline{Learning-based Component for \method.}}
We can train a model (e.g., decision tree) to predict the type of interaction for a combination. But this requires a lot more data for training (e.g., features covering various defenses and their combinations), than what we can obtain from the eight combinations in prior work. For the current defenses, our heuristic was sufficient to get a reasonable accuracy. As future work, adding a learning-based component is an interesting direction.}

\noindent\textbf{\underline{Other Causes Underlying Conflicts.}}
While we identify two possible reasons underlying conflicts among defenses (\Snospace\ref{sec:factorsInteractions}), we do not claim this to be complete.
There could be other underlying reasons which can be included in \method to make it more accurate.
One possible reasons could be the a choice of different $l_p$-norm distances for some defenses (e.g., \defevasionIn, and adversarial example-based \defmodelwmPre and \defdatawmPre). Prior works have shown that the objectives of obtaining robustness to different $l_p$-norm bounds are conflicting~\citep{lpcohrence}. 
In case of combinations, \citet{thakkar2023elevating} show that choosing different amount of noise for watermarking and adversarial training can remove a conflict. We leave the exploration of additional reasons underlying conflicts as future work.

\noindent\textbf{\underline{Other Combination Techniques.}} 
\citet{duddu2023sok} (Table 3) systematize unintended interactions among defenses and risks, categorizing them as increasing, decreasing, or unexplored. 
An alternative na\"ive technique could reject combinations where one defense increases the risks mitigated by another.
However, this is restrictive and discards several non-conflicting combinations (e.g., \defexplPost and \defmodelwm, \defevasionIn and \deffngprntPost).
Since there are several unexplored interactions in their systematization, it is challenging to applying this na\"ive technique in our context.
Hence, this technique is limited to some combinations, and not applicable to all combinations in the current state.

\change{\noindent\textbf{\underline{Other Dataset Modalities and Models.}} We rely on two image datasets previously used by prior work to evaluate most of our defenses, making them a natural starting point. Since \method’s steps are modality-independent, we conjecture that it can be applied similarly to other data types.
Evaluating \method on new modalities requires implementing corresponding defenses and models (e.g., language models for text). Then, our methodology can be followed: (a) identifying various risks and corresponding defenses in different stages of the ML pipeline, and (b) use the empirical evaluation of combinations as ground truth to verify the accuracy of \method. This is a substantial undertaking, and hence, left as future work since it may bring out new insights.
We tried to extend our existing defense implementations to tabular dataset, and indeed, not all defenses transfer to other data modalities. For example, when adapting our defenses to the CENSUS dataset, only 4 of the 11 implemented defenses were applicable—image-specific techniques like poisoning and watermarking do not apply to tabular data. This gave two valid combinations: (i) evasion robustness + explanations and (ii) group fairness + explanations (others were excluded due to incompatibility).
Both combinations aligned with \method’s predictions.
}

\noindent\textbf{\underline{Speculating Combinations with Omitted Defenses.}}
We \emph{speculate} on the omitted defense combinations from \Snospace\ref{sec:setup}: \defevasionPre, \defdiffprivPre, \deffairPre, and \deffairPost. Since \defevasionPre targets adversarial examples and makes local changes to $\dtrain$, we expect its combination with other defenses to behave similar to \defmodelwmPre and \defdatawmPre.
\defdiffprivPre and \deffairPre make global changes by transforming all data records in $\dtrain$ and should be applied before other defenses, as we expect them to avoid conflicts.
\deffairPost makes global changes in post-training stage to all predictions, and the behavior is likely to be similar to \defoutremPost which also makes global changes to $\model$ in post-training stage.
Validating these interactions is left for future work.
\change{We indicate some steps for future work to validate our speculation. For defenses omitted due to
\begin{enumerate*}[label={(\roman*)}]
    \item poor effectiveness: further work is required to design more effective defense variants (e.g., pre-training or post-training evasion robustness).
    \item incompatible datasets: set up experiments for the specific dataset where the defenses work well (e.g., tabular datasets instead of image datasets) and evaluate their combinations.
\end{enumerate*}
We can then combine with other defenses, and evaluate the combinations. We will release our code to combine new defenses with our existing ones.}

\change{\noindent\textbf{\underline{Trade-offs among Defenses.}} The current version of \method only outputs alignment and conflict and does not capture trade-offs. It is not clear how to compare the extent of gains/losses for different defenses, and there is no uniform metric that works across all defenses. For instance, 5\% loss in one defense may be much worse than 10\% loss in another, but there is no clear consensus on which is better. This may be application dependent and has to be determined a practitioner. Hence, we chose to leave this as future work.}

\noindent\textbf{\underline{Summary.}}
ML models must be protected against multiple risks simultaneously, requiring effective combination of defenses.
We systematize prior work, identify unexplored combinations, and evaluate limitations of prior techniques.
Using insights from our systematization, we present a technique, \method, which is more accurate than prior work, does not require modifying defenses, scales to more than two defenses, and applies to various defenses.

\section*{Ethical Considerations}

We use public datasets and implementations and none of our experiments require IRB approval. 

\change{\section*{Broader Impact}

Protecting ML models from various risks simultaneously is an important problem, especially in high-stakes domains where failures can have serious societal consequences. Our work advances the field of designing trustworthy ML systems by moving beyond individual risk mitigation—common in current research—to developing techniques to protect against multiple risks. We do not directly address “accountability” in this work, other than providing a way for model owners to assess effectiveness of defense combinations before deploying them. But it can be combined with additional mechanisms designed to ensure accountability in ML pipelines (such as \citet{duddu2024laminator}) for responsible deployment of defenses.}

\section*{Acknowledgments}
\noindent This work is supported in part by Intel (in the context of Private AI consortium), and the Government of Ontario (RE011-038). Vasisht is supported by IBM PhD fellowship, David R. Cheriton Scholarship, and Mastercard Cybersecurity and Privacy Excellence Graduate Scholarship. Views expressed in the paper are those of the authors and do not necessarily reflect the position of the funding agencies.
We thank Jian Liu (Zheijang University), Cong Wang (City University of Hong Kong), and Sebastian Szyller (Intel Labs) for fruitful discussions on this topic.

\bibliography{paperL}
\bibliographystyle{tmlr}

\appendix

\section{Summary of Defenses and Notations}\label{app:notations}

We summarize the different defenses and their impact on \phiacc from \Snospace\ref{sec:defenses} in Table~\ref{tab:defenses}.
\setlength\tabcolsep{0.5pt}
\begin{table}[!ht]
\caption{\textbf{Summary of defenses.} (Column \phiacc indicates impact on utility: ``\downmark'' (decrease), ``\samemark'' (no effect), ``\upmark'' $\rightarrow$ (increase).) 
}
\centering
\footnotesize
\resizebox{\columnwidth}{!}{
\begin{tabular}{ l|c|p{12.5cm} } 
\bottomrule

\toprule
 \textbf{Defense} & \phiacc & \textbf{References} \\ 
 \midrule
 \multicolumn{3}{l}{\textbf{\defevasion (Evasion Robustness)}}\\ 
 \textbullet\xspace\defevasionPre  (Data Augmentation) & \upmark & \cite{cutmix,zhang2018mixup,devries2017cutout,madry2018towards,rebuffi2021data}\\
 \textbullet\xspace\defevasionIn  (Adversarial Training) & \downmark & \cite{trades,cohen19c,lecuyer2019certified} \\
 \multirow{2}{*}{\textbullet\xspace\defevasionPost  (Input Processing)} & \downmark & \cite{nie2022DiffPure,song2018pixeldefend,guo2018countering,Das2017KeepingTB}\\
 & \samemark & \cite{grosse2017statistical,buckman2018thermometer}\\
 \midrule
 \multicolumn{3}{l}{\textbf{\defoutrem (Outlier Robustness)}}\\ 
 \textbullet\xspace\defoutremPre  (Data Augmentation) & \downmark & \cite{borgnia2021strong,DeepSweep,cretu2008,paudice2018detection,paudice2019label,jia2021scalability,jia2019efficient}\\
  \textbullet\xspace\defoutremIn  (Fine-tuning) & \samemark & \cite{diakonikolas19a,xu2019l_dmi,liu2020peer,patrini2017making,zhu2023neural,liu2018fine,wu2021adversarialBackdoor,distillationDefense}\\
 \textbullet\xspace\defoutremPost  (Pruning) & \downmark & \cite{zheng2022preactivation,zheng2022data,li2023reconstructive}\\ 
 \midrule
 \multicolumn{3}{l}{\textbf{\defmodelwm (Watermarking-M)}}\\  
 \textbullet\xspace\defmodelwmPre  (Backdoors) & \samemark & \cite{adiPaper,zhangWM,jiaEntabgled,uchida}\\
 \textbullet\xspace\defmodelwmIn  (Optimization)  & \downmark & \cite{bansal2022certified} \\
 \textbullet\xspace\defmodelwmPost  (API-based) & \samemark & \cite{dawn}\\ 
 \midrule
 \multicolumn{3}{l}{\textbf{\deffngprnt (Fingerprinting)}} \\ 
 \textbullet\xspace\deffngprntPost  (Fingerprints) & \samemark & \cite{caoFingerprint,peng2022fingerprinting,lukas2021deep,paramfingerprint,maini2021dataset}\\
 \midrule
 \multicolumn{3}{l}{\textbf{\defdatawm (Watermarking-D)}}\\
 \textbullet\xspace\defdatawmPre  (Backdoors) & \samemark & \cite{tekgul2022effectiveness,sablayrolles2020radioactive,liu2022miafingerprint}\\
 \midrule
 \textbf{\defdiffpriv (Differential Privacy)} & & \\ 
  \textbullet\xspace\defdiffprivPre  (Private Data) & \downmark & \cite{Xie2018DifferentiallyPG,torkzadehmahani2019dp}\\
 \textbullet\xspace\defdiffprivIn  (DPSGD) & \downmark & \cite{abadi2016deep,papernot2017semisupervised}\\
 \midrule
 \textbf{\deffair (Group Fairness)} & & \\  
  \textbullet\xspace\deffairPre  (Fair Data) & \downmark & \cite{datapreproc,optpreproc,lfr,feldman2015certifying}\\
 \textbullet\xspace\deffairIn  (Regularization) & \downmark &  \cite{celis2019classification,Gerrymandering,kearns2019empirical,agarwal2019fair,agarwal2018reductions,zhang2018mitigating,prejremover}\\
 \textbullet\xspace\deffairPost  (Calibration) & \downmark & \cite{pleiss2017fairness,hardt2016equality,discClassification,geyik2019fairness}\\
 \midrule
 \multicolumn{3}{l}{\textbf{\defexpl (Explanations)}} \\  
 \textbullet\xspace\defexplPost (Attributions) & \samemark & \cite{ismail2021improving,Smilkov2017SmoothGradRN,intgrad,koh17a,Wachter2017CounterfactualEW,gradcam,kim2018interpretability}\\
\bottomrule

\toprule
\end{tabular}
}
\label{tab:defenses}
\end{table}

\section{Formal Analysis of \method}\label{app:formal}

\change{\noindent A defense $\defense$ is defined as a tuple $D := (S,C,R,P)$, where:

\begin{itemize}[leftmargin=*]
    \item $S$ is the stage where $D$ is applied
    \item $C$ is the type of changes
    \item $R$ is the risk it uses
    \item $P$ is the protection scope of D, and $\emptyset \notin P$.
\end{itemize}

\noindent Let $\mathcal{D}$ be the set of all such defense. 
The input set for \method is defined as:
$X := \{(\mathbf{\texttt{D}}_1, \mathbf{\texttt{D}}_2) | \mathbf{\texttt{D}}_1 \neq \mathbf{\texttt{D}}_2,\ and\ \mathbf{\texttt{D}}_1, \mathbf{\texttt{D}}_2 \in \mathcal{D}\}$.
\method can be defined as a function:
$f: X \to \{0, 1\}$
where $f(\mathbf{\texttt{D}}_1, \mathbf{\texttt{D}}_2) = 0$ indicates conflict and $f(\mathbf{\texttt{D}}_1, \mathbf{\texttt{D}}_2) = 1$ indicates alignment.
Given the above formal model, we will discuss consistency, soundness, and completeness. 

\noindent\textbf{Consistency} means that \method never produces contradictory outputs for the same input pair.

\newtheorem{claim}{Claim}
\begin{claim}
There is no pair $ (\mathbf{\texttt{D}}_1, \mathbf{\texttt{D}}_2) $ for which \method simultaneously classifies both conflict and alignment, i.e. \method is consistent.
\end{claim}

\begin{proof}
\noindent According to the definition of $f$,
There are two rules or conflict cases:
\begin{itemize}[leftmargin=*]
    \item (c.1) If $ S_1 = S_2 $ and $ C_1 = Global $ and $ C_2 = Global $, then $ f(\mathbf{\texttt{D}}_1, \mathbf{\texttt{D}}_2) = 0 $.
    \item (c.2) If $ S_1 \neq S_2 $ and $R_1 \in P_2$, then $ f(\mathbf{\texttt{D}}_1, \mathbf{\texttt{D}}_2) = 0 $.
\end{itemize}

\noindent For alignment case: (a.1) If none of the conflict conditions hold, then $ f(\mathbf{\texttt{D}}_1, \mathbf{\texttt{D}}_2) = 1 $.

\noindent These conditions partition the input space $ X $ into disjoint sets: conflict or alignment, with no overlap. Assume that there exists $X_i$ such that $f(X_i)=0$ and $f(X_i)=1$. Since $f(X_i)=1$, according to (a.1), (c.1) and (c.2) do not hold, leading to a contradiction.
Therefore, \method never produces contradictory classifications for the same input and is consistent.
\end{proof}

\noindent\textbf{Soundness} ensures that if \method predicts a combination as effective (aligned), it should indeed be effective in practice. 

\noindent This assumption is supported by our empirical evaluation, where \method achieves:
\begin{itemize}[leftmargin=*]
    \item 90\% accuracy on previously studied defense combinations,
    \item 81\% accuracy on novel, unexplored combinations.
\end{itemize}

\noindent The low rates of false positives and false negatives in these evaluations suggest that \method can reliably predict combination effectiveness, providing practical evidence for its soundness.
A perfectly sound technique would rely on comprehensive empirical evaluation which we want to avoid by proposing an easy-to-use (approximate) technique, \method. Therefore, the soundness of \method is conditional on the assumption that these rules perfectly represent all conflict and alignment scenarios: (c.1) (c.2) and (a.1). 
However, since real-world defense interactions can be complex, involving subtle dependencies and emergent behaviors, \method may not fully captured reasons for conflict or alignment. Hence, \method is not perfectly sound and is likely to incur some errors.

\noindent\textbf{Completeness} suggests that \method can classify \emph{every possible defense pair} correctly as either conflict or alignment.
Similar to soundness, completeness is conditional on the assumption that the classification rules fully capture all conflict and alignment cases.
As a partial theoretical guarantee, we can prove that \method can classify every defense pair in its input domain, i.e., it produces a classification for every pair without leaving any case unclassified.
Assuming that the input set $X$ includes all possible distinct defense pairs, \method partially satisfies completeness by design. Formally,

\begin{claim}
    For every pair $(\text{\defenseone}, \text{\defensetwo}) \in X $, \method's function $ f $ produces a classification $ f(\text{\defenseone}, \text{\defensetwo}) \in \{0,1\} $ that identifies whether the defenses conflict or align, i.e., no conflict or alignment case is left unclassified. Therefore, if $X$ includes all distinct combinations of defenses, \method is partially complete.
\end{claim}

\begin{proof}
By construction, \method’s classification rules partition the input space 
$X$ exhaustively and exclusively by (c.1) (c.2) and (a.1).
These rules cover all possible defense pairs in  $X$ without overlap or ambiguity, ensuring that every pair is assigned a unique classification, \method is partially complete on $X$.
\end{proof}

\noindent The assumption that $X$ includes all possible defense combinations is supported by the fact that all surveyed defense methods can be represented within the \method framework.
}

\section{Formal Analysis for Multi-way Combination Algorithm}\label{app:formalMultiway}

\change{
\noindent The algorithm $F$ is for multi-way combinations and iterates through all permutations of defenses, and uses \method to check for pairwise conflicts:

\noindent (m.1) For defenses $\mathbf{\texttt{D}}_i \in \mathcal{D}$, $i=1,...,n$, $F(\mathbf{\texttt{D}}_1,..., \mathbf{\texttt{D}}_n)=0$ iff $\exists \mathbf{\texttt{D}}_i,\mathbf{\texttt{D}}_j \in \mathbf{\texttt{D}},\ f(\mathbf{\texttt{D}}_i, \mathbf{\texttt{D}}_j)=0$.

\noindent\textbf{Consistency.} According to the consistency of $f$, assume that there exists $\mathbf{\texttt{D}}_i,i=1,..., n$, s.t. $F(\mathbf{\texttt{D}}_1,...,\mathbf{\texttt{D}}_n)=0$ and $F(\mathbf{\texttt{D}}_1,...,\mathbf{\texttt{D}}_n)=1$. Since $F(\mathbf{\texttt{D}}_1,...,\mathbf{\texttt{D}}_n)=1$, $\forall \mathbf{\texttt{D}}_i,\mathbf{\texttt{D}}_j$, we have  $ f(\mathbf{\texttt{D}_i}, \mathbf{\texttt{D}}_j)=1$, leading to a contradiction. Therefore, $F$ is also consistent.

\noindent\textbf{Soundness and Completeness.} Under the assumption that $f$ is sound and complete, and considering that we take all possible combinations of defenses (as there is no limit on the input defense set of $F$), the soundness and completeness of $F$ is conditional on the assumption that (m.1) fully captures all conflict and alignment cases.

\noindent\textit{Soundness.} Since $f$ has some errors and is not perfectly sound, multi-way combinations can indeed introduce conflicts, which will result in false positives or negatives. Hence, the soundness of of multi-way \method requires comprehensive evaluation similar to pairwise evaluation.
In our evaluation, we only want to show that effective multi-way combinations—previously unexplored—are possible. 
A comprehensive evaluation to check for soundness is left as future work.}

\end{document}